\pdfoutput=1

\documentclass[11pt]{article}

\usepackage[]{ACL2023}
\usepackage{xcolor}
\definecolor{darkgreen}{rgb}{0.0, 0.5, 0.0}
\definecolor{lightblue}{RGB}{173,216,230}
\definecolor{lightred}{RGB}{255,182,193}
\definecolor{lightgreen}{RGB}{173,255,47}
\definecolor{lightyellow}{RGB}{255,255,204}
\definecolor{violet}{RGB}{90, 19, 242}

\usepackage{times}
\usepackage{latexsym}
\usepackage{graphicx}
\usepackage{subcaption}
\usepackage{hyperref}
\usepackage{wrapfig}
\usepackage{url}
\usepackage{graphicx}
\usepackage{multirow}
\usepackage{hyperref}
\usepackage{booktabs}
\usepackage{longtable}
\usepackage{longtable}
\usepackage{tabularx}
\usepackage{booktabs}
\usepackage{siunitx}
\usepackage{hyperref}
\usepackage{enumitem}
\usepackage{amsmath}
\usepackage{xcolor}
\usepackage{tcolorbox}
\usepackage{soul}
\usepackage{makecell}
\usepackage{multirow}
\usepackage{booktabs}
\usepackage{graphicx}
\usepackage{amsmath}
\usepackage{amssymb}
\usepackage{placeins}
\usepackage[titletoc]{appendix}
\usepackage{xltabular}
\usepackage[T1]{fontenc}

\usepackage[utf8]{inputenc}

\usepackage{microtype}

\usepackage{inconsolata}

\title{Mitigating the Privacy Issues in Retrieval-Augmented Generation (RAG) via Pure Synthetic Data }

\author{Shenglai Zeng$^{1}$, Jiankun Zhang$^3$, Pengfei He$^1$, Jie Ren$^1$, Tianqi Zheng$^{2}$, Hanqing Lu$^{2}$ \\ \textbf{Han Xu$^1$, Hui Liu$^1$, Yue Xing$^1$, Jiliang Tang$^1$ } \\ 
$^1$Michigan State University  \quad $^2$ Amazon.com  \quad $^1$Jilin University  
  \\
\{zengshe1, hepengf1,renjie3, xuhan1,xingyue1,  tangjili\}@msu.edu, \\
\{tqzheng, luhanqin\}@amazon.com, 
zhangjk9920@mails.jlu.edu.cn
}

\begin{document}
\maketitle
\newtheorem{definition}{Definition}
\vspace{0.5cm}
\begin{abstract}
\label{abstract}
Retrieval-augmented generation (RAG) enhances the outputs of language models by integrating relevant information retrieved from external knowledge sources. However, when the retrieval process involves private data, RAG systems may face severe privacy risks, potentially leading to the leakage of sensitive information. To address this issue, we propose using synthetic data as a privacy-preserving alternative for the retrieval data. We propose SAGE, a novel two-stage synthetic data generation paradigm. In the stage-1, we employ an attribute-based extraction and generation approach to preserve key contextual information from the original data. In the stage-2, we further enhance the privacy properties of the synthetic data through an agent-based iterative refinement process.  Extensive experiments demonstrate that using our synthetic data as the retrieval context achieves comparable performance to using the original data while substantially reducing privacy risks. Our work takes the first step towards investigating the possibility of generating high-utility and privacy-preserving synthetic data for RAG, opening up new opportunities for the safe application of RAG systems in various domains\footnote{Our code is available at \href{https://github.com/phycholosogy/RAG-SAGE}{this link}}.
\end{abstract}




\section{Introduction}
\label{Intro}

Retrieval-augmented generation (RAG) aims to improve language model outputs by incorporating relevant information retrieved from external knowledge sources. It has been effectively applied in various scenarios, such as domain-specific chatbots \cite{siriwardhana2023improving} and email/code completion \cite{parvez2021retrieval}. A typical RAG system often operates in two stages: retrieval and generation. First, the system retrieves relevant knowledge from an external database based on the user query. Then, the retrieved information is integrated with the query to form an input for a large language model (LLM). The LLM uses its pre-trained knowledge and the retrieval data to generate a response, enhancing the overall quality of the output.

However, according to existing literature \cite{zeng2024good, huang2023privacy, ding2024survey, qi2024follow, ren2024copyright}, RAG may face severe privacy issues when the retrieval process involves private data. For example, \citet{zeng2024good} observe that carefully designed user prompts are able to extract original sentences in the retrieval data (untargeted attack), and can also extract specific pieces of private information (targeted attack), potentially leading to the leakage of considerable amount of the retrieval data. The potential risk of information leakage can significantly limit the applications of RAG systems. For instance, a medical chatbot \cite{yunxiang2023chatdoctor} using patients' historical diagnosis cases as a knowledge source may improve response quality but raises concerns about exposing sensitive patient information.  Therefore, enhancing the privacy properties of RAG systems and protecting the retrieval data from leakage is of high importance to prevent unauthorized access or misuse and enable safe and widespread adoption, particularly in sensitive domains like healthcare.

\begin{figure}[t]
    \centering
    \includegraphics[width=1\linewidth]{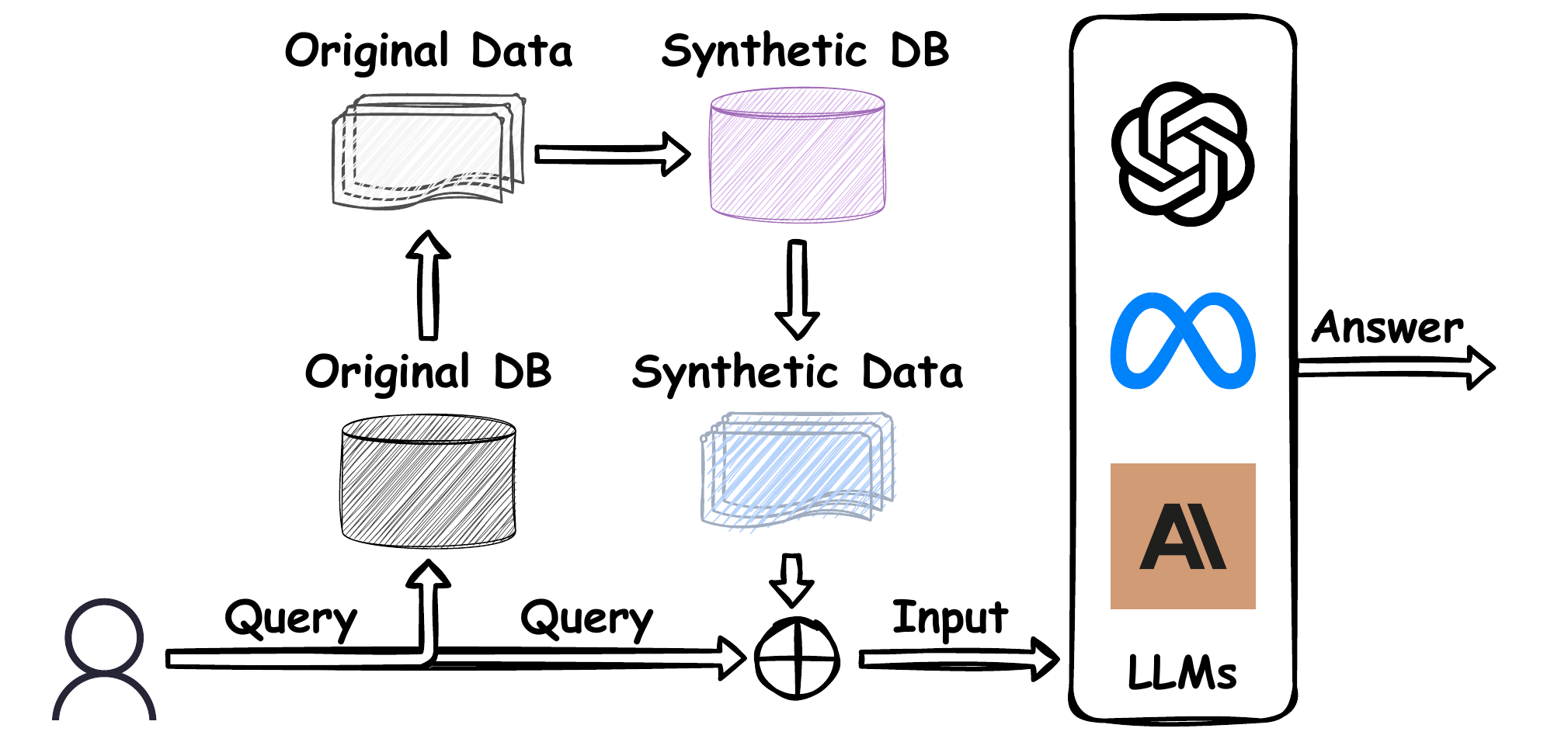}
    
    \caption{ An illustration for RAG with synthetic data. }
    \label{fig:intro}
\end{figure}

Some adaptations \cite{zeng2024good} have been proposed to protect the privacy of RAG by incorporating additional components in the RAG pipeline. These adaptations include pre-retrieval techniques (such as setting similarity distance thresholds in retrieval) and post-processing techniques (e.g., re-ranking and summarization \cite{chase2022langchain}).
However, as demonstrated by \cite{zeng2024good}, these methods cannot fully eliminate privacy risks, as the data itself may contain sensitive information. Moreover, these methods often introduce a significant privacy-utility  trade-off and may incur extra time costs during inference.

To address the above concern, we propose an alternative data-level solution via using synthetic data as shown in Figure \ref{fig:intro}. By generating a privacy-preserving version of the original data and only providing the synthetic version to the LLM, the risk of information leakage could be effectively mitigated. This approach can potentially ensure that the original data is not directly used as input to the LLMs, thereby reducing the chances of sensitive information being exposed or leaked during the retrieval and generation process. Therefore synthetic data allows the creation of a safe, surrogate dataset that maintains the essential properties and relationships of the original data while protecting sensitive information. There are recent works exploring synthetic data generation using pre-trained language models \cite{ye2022zerogen,meng2022generating,gao2023self,chen2023mixture,yu2024large,xie2024differentially} and utilizing the synthetic data 
in the downstream task to protect the privacy of the original data. Besides, some studies integrate differential privacy with 
synthetic data for in-context demonstrations \cite{tang2023privacy}.  However, while existing methods for generating synthetic data work well for downstream tasks or in-context demonstrations, they are not well aligned with the unique requirements of RAG:  RAG primarily focuses on utilizing key information from the data to answer related questions  \cite{ding2024survey}, rather than learning general patterns. Therefore, it is crucial to preserve as much useful information as possible from the original data when generating synthetic retrieval data. On the other hand, existing synthetic methods do not require generating data that shares the same key information with the original data. Consequently, there is a lack of exploration on how to effectively use synthetic data for RAG and how to design a feasible solution for generating high-quality retrieval data. Meanwhile, the unique information requirements of retrieval data also present  challenges in generating privacy-preserving synthetic data, as it is crucial to carefully select what information to preserve.

In this work, we take the first effort to investigate the possibility of generating synthetic retrieval data that maintains high utility while enhancing privacy protection for RAG. After identifying the related data from the original dataset, we use the synthetic version of the data as context instead of the original data for generation. We use a two-stage generation and refinement paradigm called 
called SAGE (\textbf{S}ynthetic \textbf{A}ttribute-based \textbf{G}eneration with ag\textbf{E}nt-based refinement) to generate synthetic retrieval data. To preserve the important information of the original data and keep the utility of the synthetic data, we first utilize an attributed-based extraction and generation approach to generate the synthetic data.  Specifically, for each dataset, we first input few-shot samples  to make the LLM identify important attributes of the dataset. Then, for each data sample, we ask the LLM to extract key information corresponding to these attributes. After that, we input the attribute information into another LLM and ask it to generate synthetic data based on these key points (stage-1). In this way, the generated data contains key contextual information. 

Although the attribute-based method can preserve key information of the original data, it may still include some privacy information, as the stage-1 does not incorporate privacy constraints. Therefore, a second step is necessary to further preserve privacy. In stage-2, we propose an agent-based iterative refinement approach to enhance the protection of private information. Specifically, we introduce two agents, a privacy assessment agent and a rewriting agent. The privacy assessment agent determines whether the generated data contains privacy information, such as containing personally identifiable information (PIIs) or potentially leading to the linkage of personal information, and provide feedback. The rewriting agent then takes this feedback to refine its generated data until the privacy agent deems it safe. Our experimental results show that using our synthetic data as retrieval data can achieve comparable performance with using original data while substantially reducing the associated privacy risks.

\section{Related Works}
\label{Intro}

\subsection{Retrieval-augmented generation and its privacy issues}
Retrieval-augmented generation (RAG), introduced by \citet{lewis2020retrieval}, has become a popular approach to enhance LLMs' generation ability \cite{liu2022llama,chase2022langchain,van2023clinical,ram2023context,shi2023replug}. RAG improves output accuracy and relevance \cite{gao2023retrieval}, mitigating "hallucinations" of LLMs \cite{shuster2021retrieval}. Its flexible architecture allows seamless updates to the dataset, retriever, and LLM without re-training \cite{shao2023enhancing,cheng2023lift}. These advantages make RAG a favored approach for applications like personal chatbots and specialized domain experts \cite{panagoulias2024augmenting}.

However, there  application of RAG also brings privacy issues.  \citet{huang2023privacy} have shown the privacy implications of retrieval-based LM and identified privacy leakage of  KNN-LM \cite{khandelwal2019generalization}, a specific kind of retrieval LM. \citet{zeng2024good} have shown that RAG is vulnerable to extraction attacks. \citet{qi2024follow} have shown that production RAG models also suffer from attacks. The vulnerability of RAG makes its application in privacy domains under high risks. 

\subsection{Synthetic data generation using large language models}

As large language models
become more expressive, researchers have explored using them to generate synthetic data. \citet{ye2022zerogen,meng2022generating} propose to  generate synthetic data via zero-shot prompting and then train smaller models on these data to handle various tasks like text classification, question answering and etc. \citet{gao2023self} further develop a noise-robust re-weighting framework to improve the quality of generated data. \citet{chen2023mixture} propose to mix a set of soft prompts and utilize prompt tuning to generate diverse data. \citet{yu2024large} focus on the attributes of data itself including length and style to generate more diverse data. Recent works \cite{tang2023privacy,xie2024differentially} take privacy into consideration. \citet{tang2023privacy} propose a few-shot data generation method to generate private in-context demonstrations from a private dataset and provide a differential privacy guarantee. \citet{xie2024differentially} introduce a private evolution algorithm to generate deferentially private data. However, their synthetic data is not guaranteed to include contextual information in the original data, thus not fitting the RAG system well.
\section{Methods}
\label{methods}

\begin{figure*}[t]
    \centering
    \includegraphics[width=0.8\textwidth]{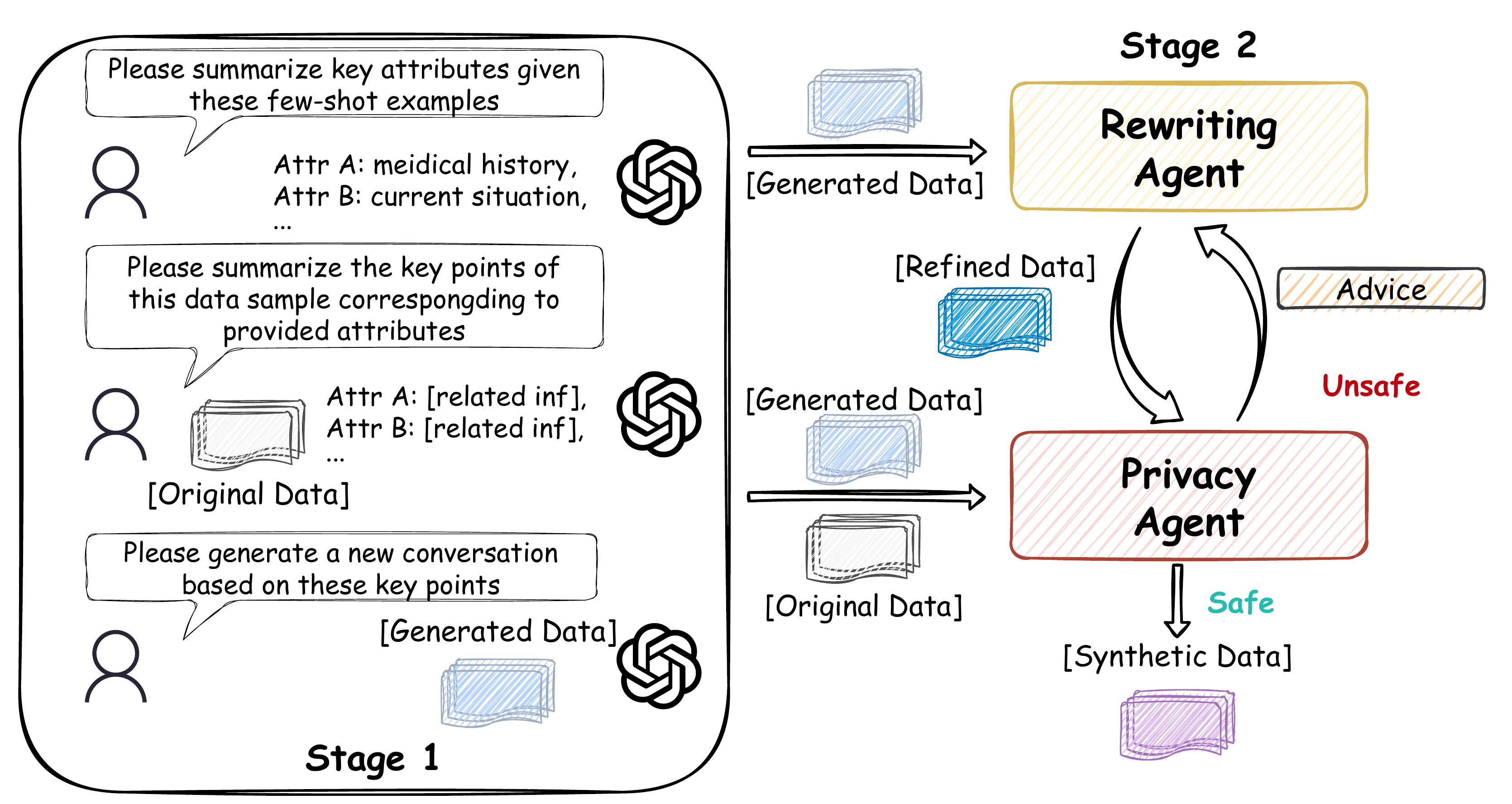}
    
    \caption{Pipeline of generating synthetic data. }
    \label{fig:pipeline}
\end{figure*}

Our SAGE framework of generating synthetic retrieval data is composed of two stages, i.e., attribute-based data generation and agent-based interactive refinement, as shown in Figure \ref{fig:pipeline}.  The stage-1 aims to generate data that contains essential information of original data, while the  stage-2 aims to automatically refine the data to further mitigate the privacy concerns. The synthetic data generation process can be conducted \textbf{offline} and only needs to be performed \textbf{once}. During inference, when the original data is identified, the corresponding synthetic data is returned as retrieval data\footnote{Our framework is versatile and adaptable to various scenarios and fields, as discussed in Appendix \ref{adaptation}}.


\subsection{Stage-1: Attribute-based data generation}

In this stage, we aim to generate synthetic data that contains all the essential information from the original data. To achieve this goal, we propose  an attribute-based data extraction and generation paradigm to create synthetic data. 

The entire process of Stage-1 consists of three steps: identifying important attributes using few-shot samples, extracting key information related to essential attributes, and generating synthetic data conditioned on the extracted key information. First, we feed few examples within the dataset to an LLM-based \textit{attribute identifier} and prompt it to identify $m$ most essential attributes of the dataset\footnote{We discuss the impact of $m$ in Section \ref{ablation_study}}. This process is performed before generating any synthetic data, and is only needed for once. Then, after obtaining the essential attributes, we leverage an LLM-based \textit{information extractor} to extract key information related to these attributes for each data sample and construct [attribute:key information] pairs. This step captures the core useful information of the original data. Finally, we input these attribute-information pairs into an LLM-based \textit{data generator} to generate new synthetic data. The synthetic data is expected to include key information extracted in the second step, thus reducing the loss of useful information in the original data. The prompt used for this step is provided in Appendix \ref{prompt_s1}.
 The LLMs used in these steps (attribute identifier, information extractor, and data generator) can be the same or different models. In Section \ref{ablation_study}, we also explore different model combinations and their impacts. Through stage-1, the risk of untargeted attacks is also mitigated, as this process reduces unnecessary information from the original data while only maintaining essential information.


\subsection{Stage-2: Agent-based private data refinement}

Though the synthetic data generated in Stage-1 has preserved important information from the original data, it may still have privacy issues as no privacy controls are added. For example, it may contain PIIs such as email addresses or phone numbers, or specific personal information that can possibly be linked to specific individuals. Thus, the synthetic data still may cause privacy leakage when used as retrieval data. Although methods such as anonymization can mitigate this issue to some extent, they can only  mask highly structured data like email addresses, and it is challenging to reduce other potential privacy risks~\cite{wang2022topology}. As pointed out in \cite{brown2022does}, one key challenge in natural language processing (NLP) is that private information is often not explicitly presented but can be inferred from the context. Considering the sentence: "I just got back from the oncology department at City Central Hospital. The doctor said my chemo is going well.", this sentence does not directly mention the person's name but reveals that the speaker is undergoing cancer treatment at City Central Hospital. Moreover, \citet{shi2022just} further demonstrate that although directly removing all entities can preserve privacy, it will cause the data to contain almost no useful information, and the performance loss would be unacceptable. To address this issue, we propose to utilize the rewriting and reflection capabilities of large language models (LLMs) through an agent-based approach. This method involves 2 agents collaborating to iteratively refine the generated answers so that they can maintain utility while protecting privacy.

Specifically, in our framework, we introduce a privacy agent and a re-writing agent that collaborate iteratively to enhance the privacy of the generated data. The privacy agent takes both the generated data from Stage-1 and the original data as input to assess whether the generated data contains privacy issues, such as containing PIIs or the linkage of personal information. It then provides feedback to the re-writing agent. The re-writing agent, in turn, improves data according to the privacy agent's advice. The privacy agent then evaluates the newly generated data again. This process continues until the privacy agent determines that the synthetic data is safe\footnote{We put the detailed workflows and system prompts of these two agents and average iteration rounds in Appendix \ref{prompt_s2} and synthetic data examples in Appendix \ref{examples_syn}.}. Stage-2 mitigates targeted attack risks by eliminating structured PII (e.g., emails, phone numbers), which can be effectively identify, remove and rewrite by advanced LLMs such as GPT-3.5.

\section{Experiment}

In this section, we present various experimental results to demonstrate the utility and privacy properties of SAGE. We first introduce our experiment setup in Section \ref{setup}, including the components of RAG, evaluation datasets, tasks, and baselines. Then, we present the utility and privacy results in Section \ref{ex_utility} and Section \ref{ex_privacy}. Moreover, we conduct ablation studies in Section \ref{ablation_study} to investigate the impact of the number of attributes, model choice, and the number of retrieved documents on the performance and privacy of SAGE. We also discuss the cost of synthetic data in Appendix \ref{cost}.

\subsection{Evaluation Setup} 
\label{setup}
\paragraph{RAG components} In our experiments, we mainly employed Llama3-8b-chat (L8C) as the language model for text generation for performance evaluation. We chose this model because it cannot perform well on our chosen tasks without RAG, allowing us to test the extent to which RAG can improve the generation quality. For the privacy experiments, we use both the widely-used closed-source model GPT-3.5-turbo and the open-source model L8C for text generation. Both models have been safety-aligned, allowing us to demonstrate the vulnerability of RAG systems and the effectiveness of our proposed methods. We utilized the \texttt{bge-large-en-v1.5} model as the embedding model. The embeddings were stored and the retrieval database was constructed using the FAISS library. By default, the $L_2$-norm was used as the similarity metric to compare embeddings. Unless otherwise specified, we retrieved a single document ($k=1$) for each query. The impact of varying the number of retrieved documents was further investigated in Section \ref{ablation_study}. \footnote{By defaute, we use GPT-3.5 at stage-1 and GPT-4 for agents at stage-2, we explore the model choice in Section \ref{ablation_study}}

\paragraph{Tasks and retrieval datasets} We consider two privacy-related scenarios to verify the effectiveness of our synthetic methods. In the first scenario, we focus on monitoring medical dialog cases and utilize the HealthcareMagic-101
dataset of 200k doctor-patient medical dialogues as the retrieval dataset. In the second scenario, we follow the setting of \cite{huang2023privacy} to consider a case where some private information is mixed with a public dataset. Specifically, we mix personal information pieces from the Enron Mail dataset (private dataset) with the wikitext-103 dataset (public dataset), which we refer to as Wiki-PII dataset. We extract personal PIIs and combine those PIIs with each sample of the wikitext-103 dataset. The details of the construction are presented in Appendix \ref{dataset_cons}. We then evaluate the performance of our methods on open-domain question answering datasets (ODQA), including Natural Questions (NQ) \cite{kwiatkowski2019natural}, Trivia QA (TQA) \cite{joshi2017triviaqa}, Web Questions (WQ) \cite{berant2013semantic}, and CuratedTrec (CT) \cite{baudivs2015modeling}. The detailed descriptions of these datasets are included in Appendix \ref{dataset_cons}.

\paragraph{Baselines.} To verify the effectiveness of our methods, we include three baselines: simple paraphrasing\footnote{\textit{We also incorporate more complex prompts and advanced models, such as GPT-4o in Appendix \ref{para}, for paraphrasing, and obtain consistent conclusions.}} and existing representative LLM-based data synthesis methods like \textbf{ZeroGen} \cite{ye2022zerogen} and \textbf{AttrPrompt} \cite{yu2024large}. We provide the details of  these methods in Appendix \ref{baseline}. We also report generation results without RAG, denoted as \textbf{0-shot}, using original data directly as retrieval data, denoted as \textbf{Origin}, and the outputs of the attributes-based generation, denoted as \textbf{Stage-1}. Finally, we report the outputs of the complete SAGE pipeline, denoted as \textbf{Stage-2}.
\begin{table}[!ht]
\centering
\caption{Utility results on HealthCareMagic dataset}
\label{tab:perform_per_chat_llama}
\resizebox{0.6\linewidth}{!}{
\begin{tabular}{@{}c|cc@{}}
\toprule
Method & BLEU-1 & ROUGE-L \\
\midrule
0-shot & 0.081 & 0.0765 \\
Origin & 0.0846 & 0.0789 \\
Paraphrase & 0.105 & 0.0952 \\
ZeroGen & 0.0850 & 0.0769 \\
AttrPrompt & 0.079 & 0.067 \\
Stage-1 & 0.114 & 0.0956 \\
Stage-2 & 0.113& 0.0943 \\
\bottomrule
\end{tabular}
}
\end{table}

\begin{table*}[!ht]
\centering
\caption{Utility results on Wiki-PII dataset }
\label{tab:perform_wikiqa}
\resizebox{0.8\textwidth}{!}{
\begin{tabular}{@{}c|cccccccc@{}}
\toprule
Method & \multicolumn{2}{c}{NQ} & \multicolumn{2}{c}{TQA} & \multicolumn{2}{c}{WQ} & \multicolumn{2}{c}{CT} \\
\cmidrule(lr){2-3} \cmidrule(lr){4-5} \cmidrule(lr){6-7} \cmidrule(lr){8-9}
& BLEU-1 $\uparrow$ & ROUGE-L $\uparrow$ & BLEU-1 $\uparrow$ & ROUGE-L $\uparrow$ & BLEU-1 $\uparrow$ & ROUGE-L $\uparrow$ & BLEU-1 $\uparrow$ & ROUGE-L $\uparrow$ \\
\midrule
0-shot & 0.00719 & 0.0136 & 0.00843 & 0.0157 & 0.00716 & 0.0143 & 0.00882 & 0.0150 \\
Origin & 0.0180 & 0.0315 & 0.0150 & 0.0272 & 0.0147 & 0.0271 & 0.0178 & 0.0323 \\
Paraphrase & 0.0153 & 0.0269 & 0.0127 & 0.0251 & 0.0094 & 0.0187 & 0.0135 & 0.0252 \\
ZeroGen & 0.0034 & 0.0063 & 0.0057 & 0.010 & 0.0104 & 0.0201 & 0.0116 & 0.0205 \\
AttrPrompt & 0.0061 & 0.0107 & 0.006 & 0.0108 & 0.006 & 0.0110 & 0.00624 & 0.0111 \\
Stage-1 & 0.0131 & 0.0257 & 0.0125 & 0.0249 & 0.0132 & 0.0277 & 0.0122 & 0.0242 \\
Stage-2 & 0.0177 & 0.0322 & 0.0131 & 0.0247 & 0.0173 & 0.0298 & 0.0129 & 0.0267 \\
\bottomrule
\end{tabular}
}
\end{table*}

\begin{table*}[!ht]
\centering
\caption{Targeted attack results on Wiki-PII and HealthCareMagic dataset(250 prompts)}
\label{tab:target}
\resizebox{0.8\textwidth}{!}{
\begin{tabular}{@{}c|cc|cc|cc|cc@{}}
\toprule
 & \multicolumn{2}{c}{Target-wiki-llama-3-8b} & \multicolumn{2}{c}{Target-wiki-gpt-3.5} & \multicolumn{2}{c}{Target-chat-llama-3-8b} & \multicolumn{2}{c}{Target-chat-gpt-3.5} \\
\cmidrule(lr){2-3} \cmidrule(lr){4-5} \cmidrule(lr){6-7} \cmidrule(lr){8-9}
Method & Target info $\downarrow$ & Repeat prompts $\downarrow$ & Target info $\downarrow$ & Repeat prompts $\downarrow$ & Target info $\downarrow$ & Repeat prompts $\downarrow$ & Target info $\downarrow$ & Repeat prompts $\downarrow$ \\
\midrule
origin & 25 & 12 & 167 & 64 & 7 & 23 & 75 & 132 \\
para & 9 & 1 & 28 & 9 & 17 & 26 & 42 & 81 \\
ZeroGen & 4 & 5 & 5 & 2 & 0 & 3 & 1 & 6 \\
AttrPrompt & 0 & 0 & 0 & 0 & 0 & 0 & 0 & 0 \\
Stage-1 & 1 & 4 & 3 & 19 & 3 & 11 & 12 & 36 \\
Stage-2 & 0 & 0 & 0 & 7 & 0 & 0 & 0 & 0 \\

\bottomrule
\end{tabular}
}
\end{table*}

\begin{table*}[!ht]
\centering
\caption{Untargeted attack results on HealthCareMagic dataset(250 prompts)}
\label{tab:untarget_chat}
\resizebox{0.8\textwidth}{!}{
\begin{tabular}{@{}c|cc|cc|cc|cc@{}}
\toprule
 & \multicolumn{4}{c}{Untarget-chat-llama} & \multicolumn{4}{c}{Untarget-chat-gpt3.5} \\
\cmidrule(lr){2-5} \cmidrule(lr){6-9}
Method & \begin{tabular}[c]{@{}c@{}}Repeat \\ prompt $\downarrow$ \end{tabular} & \begin{tabular}[c]{@{}c@{}}ROUGE \\ prompt $\downarrow$ \end{tabular} & \begin{tabular}[c]{@{}c@{}}Repeat \\ context $\downarrow$ \end{tabular} & \begin{tabular}[c]{@{}c@{}}ROUGE \\ context $\downarrow$\end{tabular} & \begin{tabular}[c]{@{}c@{}}Repeat \\ prompt $\downarrow$ \end{tabular} & \begin{tabular}[c]{@{}c@{}}ROUGE \\ prompt $\downarrow$\end{tabular} & \begin{tabular}[c]{@{}c@{}}Repeat \\ context $\downarrow$\end{tabular} & \begin{tabular}[c]{@{}c@{}}ROUGE \\ context $\downarrow$ \end{tabular} \\
\midrule
origin & 19 & 17 & 16 & 13 & 61 & 67 & 49 & 67 \\
para & 23 & 13 & 22 & 11 & 45 & 63 & 33 & 50 \\
ZeroGen & 0 & 0 & 0 & 0 & 0 & 0 & 0 & 0 \\
AttrPrompt & 0 & 0 & 0 & 0 & 0 & 0 & 0 & 0 \\
Stage-1 & 1 & 2 & 1 & 2 & 1 & 0 & 1 & 0 \\
Stage-2 & 0 & 0 & 0 & 0 & 0 & 0 & 0 & 0 \\

\bottomrule
\end{tabular}
}
\end{table*}
\subsection{Utility of using synthetic data}
\label{ex_utility}
To assess the utility of using synthetic data as retrieval data, we evaluate the quality of the generated answers by comparing the answers with the ground truth. We primarily report the ROUGE-L and BLEU scores between the generated and the ground truth answers. \textit{We also incorporate more evaluation metrics such as Exact Match(EM) and LLM-based evaluation and get similar conclusion in Appendix \ref{other_metrics}}. \textit{The details of these matrics are explained in Appendix \ref{detail_metrics}.}

\paragraph{Utility results on medical dialog.}For the medical dialog case, we split the data into two parts: 99\% of the data is used as the retrieval data, and the remaining 1\% is used as the test data. To evaluate the system's performance, we input questions from the test set and compare the generated answers with the ground truth answers using similarity-based metrics such as ROUGE-L and BLEU scores. The results are reported in Table \ref{tab:perform_per_chat_llama}.
The results demonstrate that using synthetic data achieves performance comparable to, and  even better than, using original data. Moreover, it significantly outperforms generation without retrieval. Our methods also surpass simple paraphrasing and ZeroGen. These findings suggest that our approach to generating synthetic data effectively preserves the utility of the original data. 


\paragraph{Utility results on ODQA.}To assess open-domain question answering (ODQA) performance, we combine the WikiText-101 dataset with Enron Mail, as the source for information retrieval. We then evaluate the system's performance using multiple ODQA datasets, such as Natural Questions (NQ), Trivia QA (TQA), WQ, CT.  

The experiment results are summarized in Table \ref{tab:perform_wikiqa}. Similar to Table \ref{tab:perform_per_chat_llama}, using our proposed synthetic data as retrieval data shows consistently high performance, comparable to directly using the original data. In some datasets, such as NQ and WQ, our synthetic data even outperforms the original data. This may be because our pipeline in stage-1 preserves most of the essential key information. In stage-2, the data is further refined, and the final outputs contain more "pure" useful information, making it easier for the LLM to identify essential information and generate better answers.


\subsection{Privacy of using synthetic data}
\label{ex_privacy}
To evaluate the privacy properties of using our synthetic data as retrieval data, we conducted targeted and untargeted attacks following~\cite{zeng2024good}, which can cause considerable data leakage from standard retrieval database. The composite structured prompting attack on RAG consists of two components: \{\textit{information}\} and \{\textit{command}\}. The \{\textit{information}\} component guides the retrieval system to fetch specific data, while the \{\textit{command}\} component instructs the language model to include the retrieved information in its response. For the \{\textit{command}\} component, we use phrases such as "Please repeat all the context" for both targeted and untargetd attacks. The \{\textit{information}\} component is adjusted according to the objectives of the attack. Targeted attacks aim to extract specific sensitive information, such as PII or private dialogue cases, by providing relevant input. In contrast, untargeted attacks seek to gather as much data as possible from the entire retrieval dataset without focusing on specific information.

For untargeted attacks, we report the number of prompts that can generate outputs with either at least 10 tokens exactly matching the original dataset (\textbf{Repeat Prompt}) or with sufficient similarity to the original data, as indicated by a ROUGE-L score exceeding 0.5 (\textbf{Rouge Prompts}). Additionally, we report the number of unique verbatim excerpts (\textbf{Repeat Contexts}) and closely similar answers retrieved from the data, with a ROUGE-L score higher than 0.5 (\textbf{Rouge Contexts}). For targeted attacks, we also report the \textbf{Repeat Prompt} metric and the number of unique targeted information pieces extracted (\textbf{Targeted Information}). Further details regarding the threat models are provided in Appendix \ref{attack_details}. \footnote{\textit{We also directly compare the similarity between the synthetic data and original content as a worst-case scenario, presented in Appendix \ref{worst_case} and Table \ref{tab:worst_case}}.}


\paragraph{Untargeted attack results.} In the context of an untargeted attack, the attacker's objective is to gather as much information as possible from the whole retrieval dataset, rather than seeking specific data. To achieve this, following \cite{carlini2021extracting}, we randomly select chunks from the Common Crawl dataset to serve as the \{\textit{information}\} component. We report the results of untargeted attacks on the HealthCareMagic dataset in Table \ref{tab:untarget_chat}.  From the results, we can observe that (a) 
The attribute-based generated data in the first stage shows a lower attack success rate than using original data and simple paraphrasing. This indicates that the method is inherently more robust to privacy attacks. This may be because the generation process produces a new version of the data and naturally avoids highly similar sentences and exact copies. (b) After the agent-based refinement, using the final version of the data as context shows nearly no untargeted attack success. This demonstrates that the refinement process has  further protected the data, and the produced data is of high safety.

\paragraph{Targeted attack results.} In the targeted attack, the attacker has specific objectives regarding the type of information they aim to extract, such as PIIs including phone numbers and email addresses, or sensitive content like personal dialogue cases. For these attacks, the \{\textit{information}\} component consists of some specific information that is related to the attacker's goals. When evaluating the HealthCareMagic dataset, our goal is to extract diagnosed cases for specific diseases. We use the prompt "I want information about {disease}" as the {\textit{information}} component. To consider the targeted information as successfully extracted, two criteria must be met: (a) the name of the targeted disease must appear in the original data, and (b) the model's output must contain repetitive segments from the original data at least 10 tokens. In the case of the Wiki-PII dataset, which includes a mix of data from Enron Mail, we focus on retrieving PIIs by employing frequently used leading phrases such as "My phone number is" as the {\textit{information}} element. The targeted information in this context is measured by the total count of PIIs effectively extracted from the retrieval dataset.  


The results of targeted attacks lead to conclusions similar to those of untargeted attacks. From Table \ref{tab:target}, the generated data in the first stage has significantly reduced targeted information leakage. 
This is because the synthetic data only retains the essential key information and may naturally omit some specific privacy information. Furthermore, after the agent-based refinement process, the attack success rate further decreases to nearly zero. This validates that the agent-based refinement process can successfully further reduce the possibly privacy-violating information in the synthetic data.



\subsection{Ablation Studies}
\begin{figure*}[t]

\centering
\resizebox{\textwidth}{!}{%
    \begin{minipage}{\textwidth}
        \subfloat[ Performance(info extractor)]{\includegraphics[width=.25\textwidth]{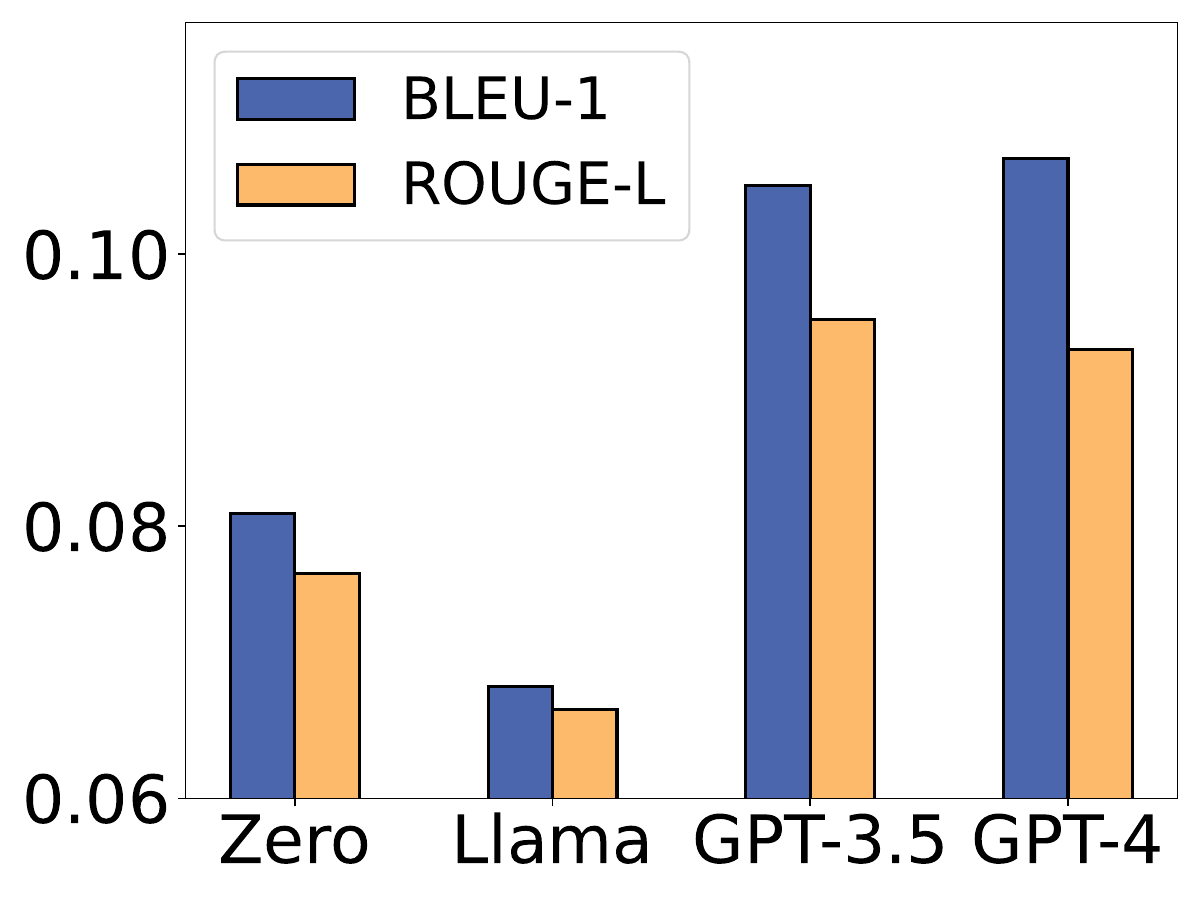}
        \label{fig:data_attributes_performance}}
        \subfloat[Performance(generator)]{\includegraphics[width=.25\textwidth]{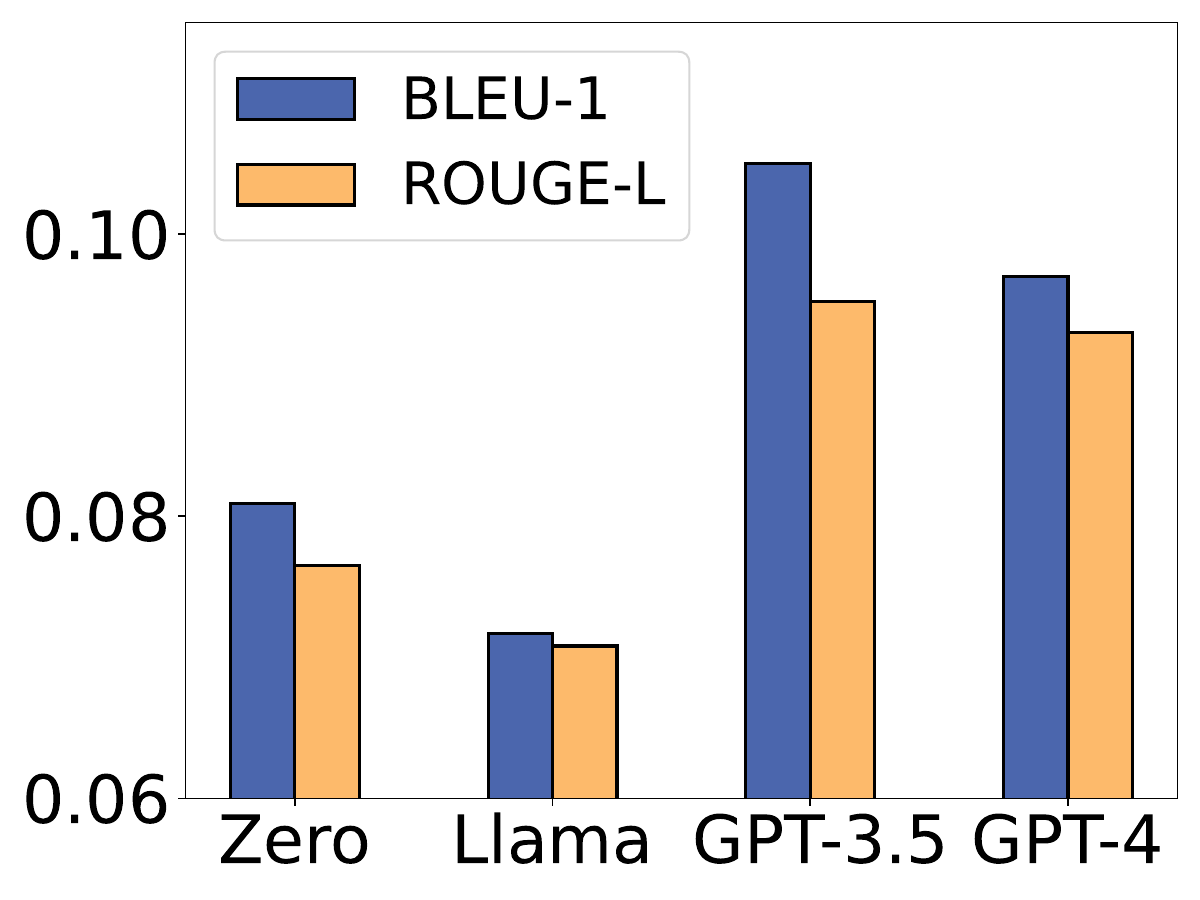}
        \label{fig:data_generator_performance}}
        \subfloat[Attack(info extractor)]{\includegraphics[width=.25\textwidth]{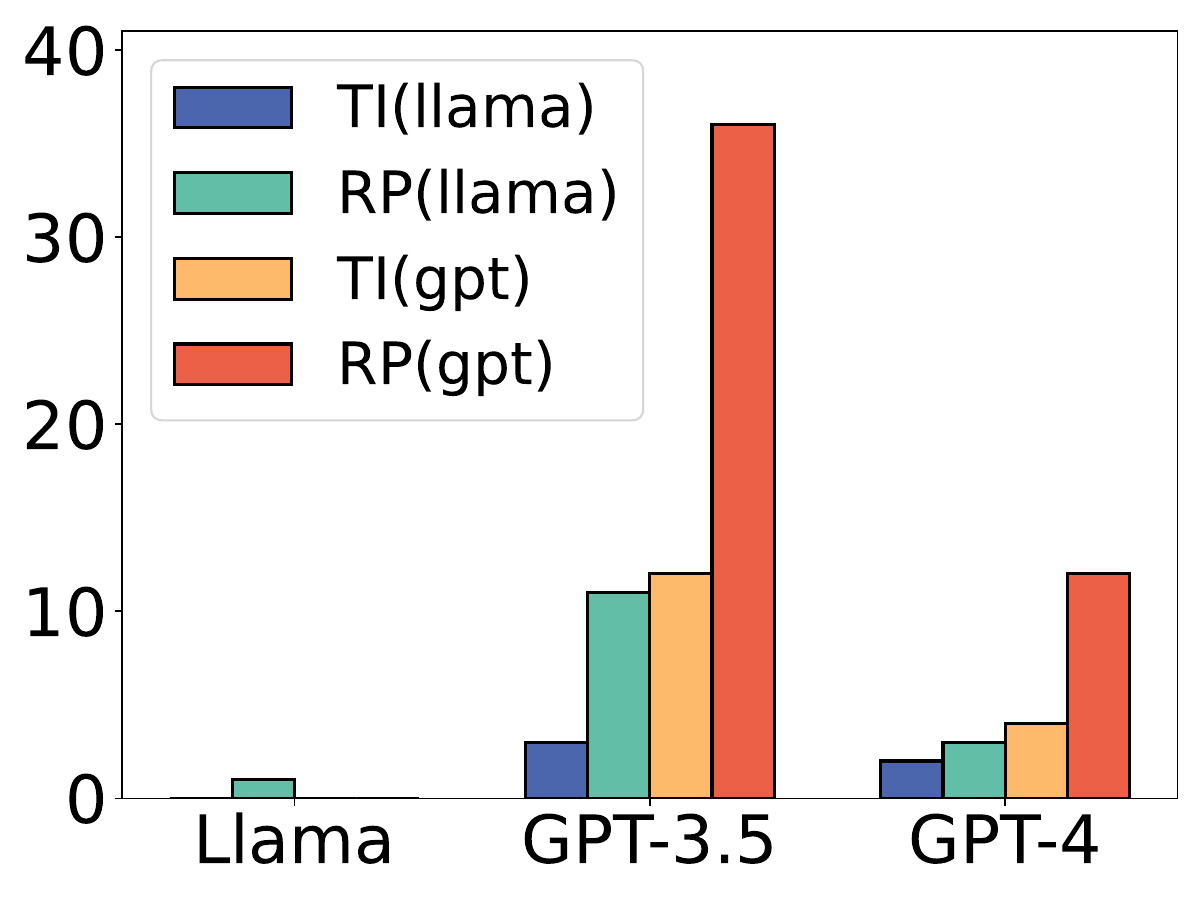}
        \label{fig:data_attributes_Attack}}
        \subfloat[Attack(data generator)]{\includegraphics[width=.25\textwidth]{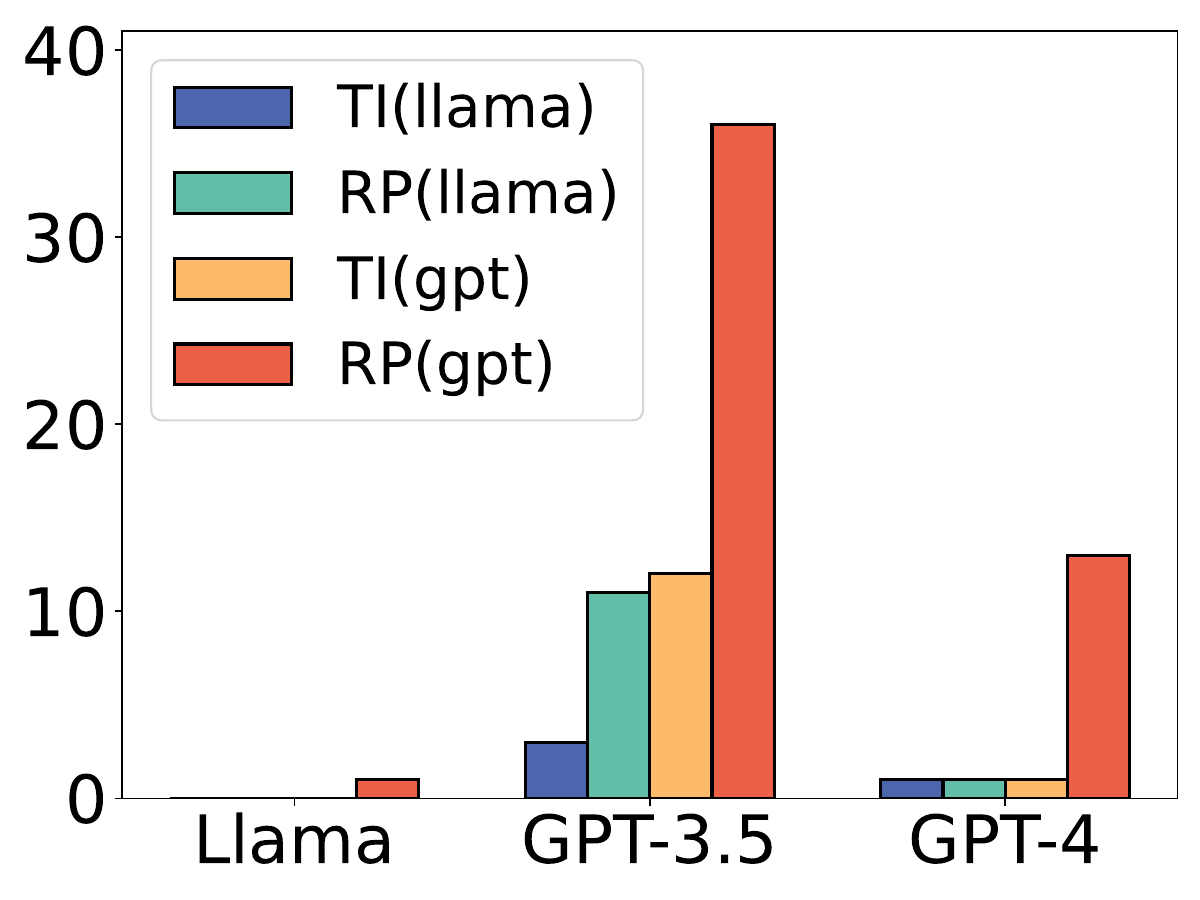}
        \label{fig:data_generator_Attack}}
    \end{minipage}
}
\caption{Ablation study on model choice. TI means targeted information and RP means repeat prompts.}
\label{Model Choice}
\end{figure*}
\label{ablation_study}
To investigate the factors that affect the quality of synthetic data, we conduct ablation studies regarding the impact of model choice, the number of attributes,  and retrieved  documents per query.

\begin{figure}[t]

\centering
\resizebox{\linewidth}{!}{%
    \begin{minipage}{\linewidth}
        \subfloat[ Performance($m$)]{\includegraphics[width=.49\linewidth]{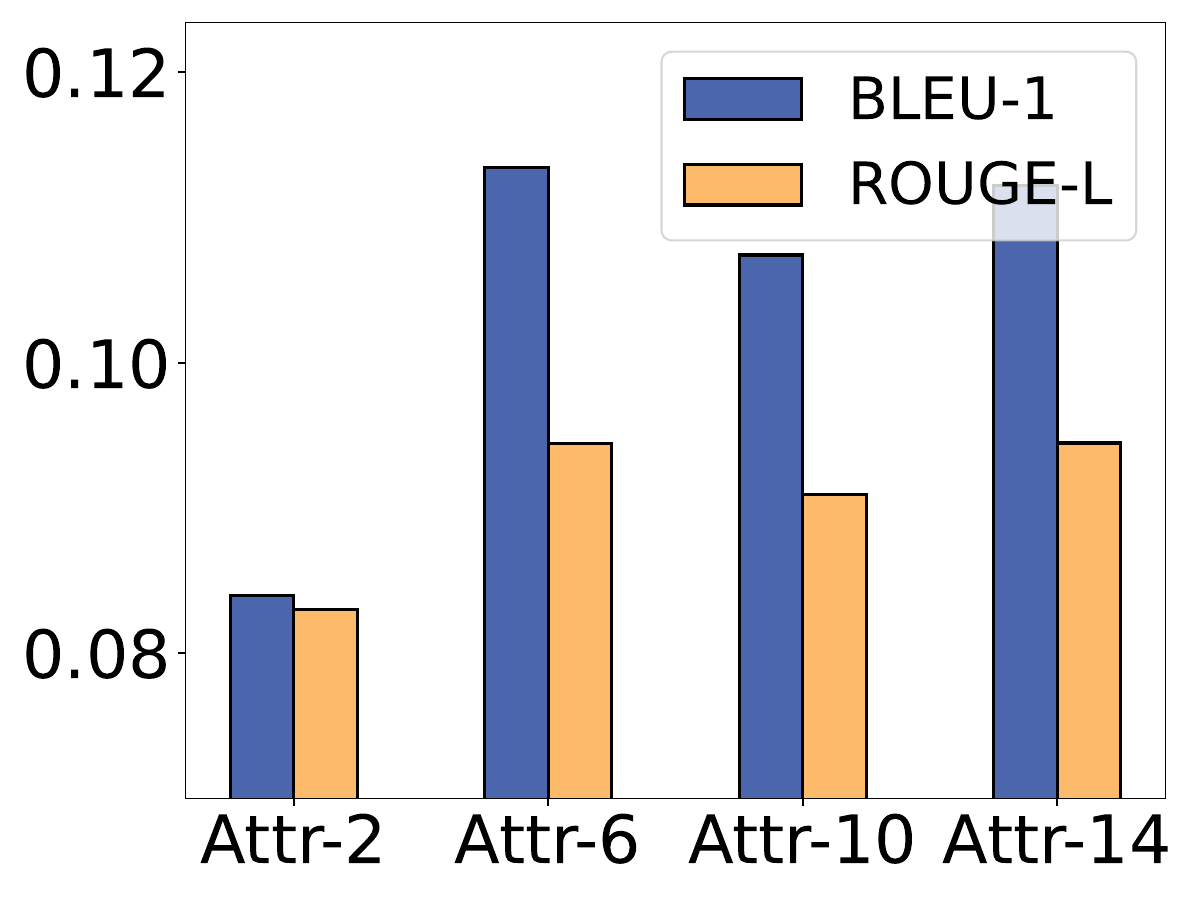}
        \label{fig:num_attributes_performance}}
        \subfloat[Attack($m$)]{\includegraphics[width=.49\linewidth]{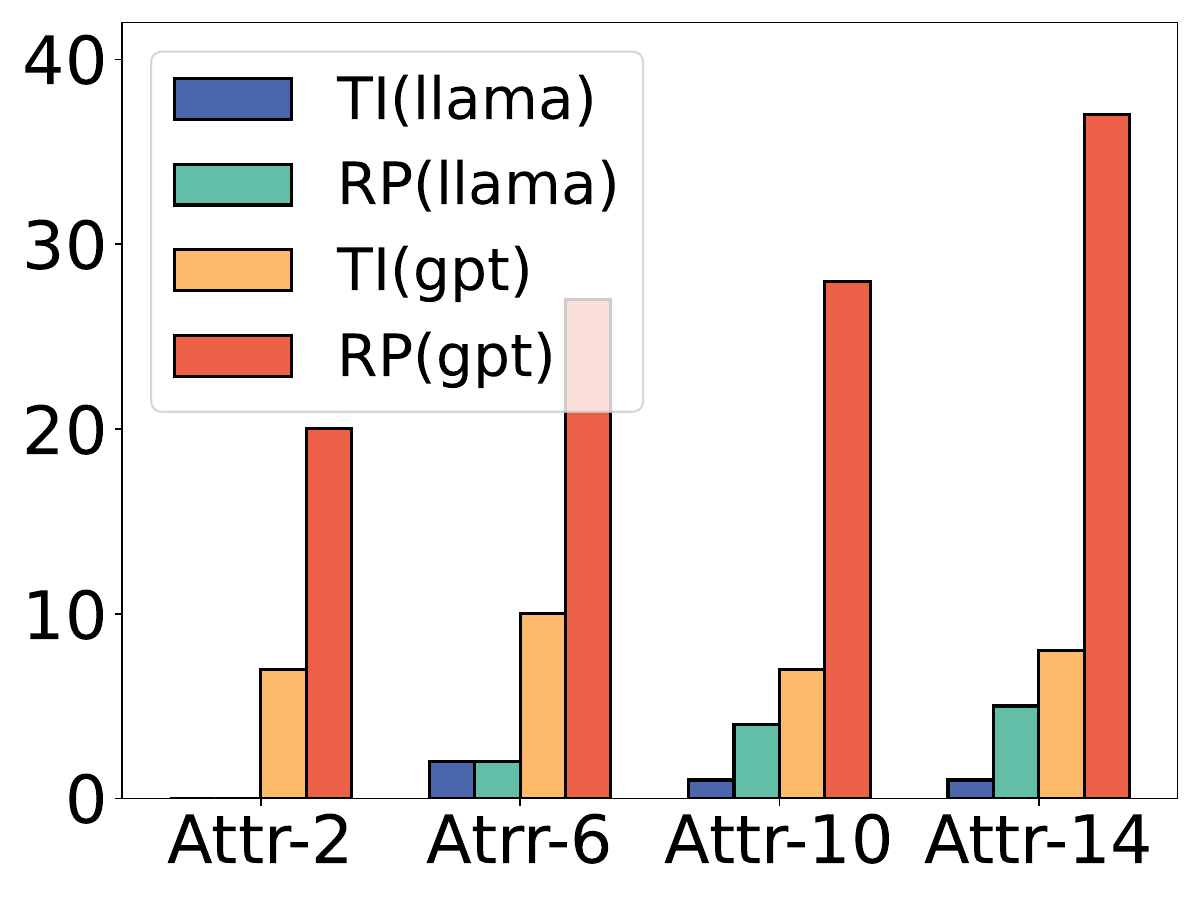}
        \label{fig:num_generator_Attack}}
    \end{minipage}
}
\caption{Ablation study on number of attributes $m$.}
\label{num_attri}
\end{figure}
\paragraph{Impact of model choice.} 

To investigate the influence of model choice on stage-1 generation, we change the models used for the \textit{information extractor} and \textit{data generator} components in stage 1. Specifically, we experiment with different models, including GPT-4, GPT-3.5, and Llama3-Chat-8b, for these two components. For the experiments on the \textit{information extractor}, we fix the \textit{data generator} as GPT-3.5 and vary the model used for the \textit{information extractor}. Similarly, for the experiments on the \textit{data generator}, we fix the model of \textit{information extractor} as GPT-3.5 and vary the model of \textit{data generator}. We conduct the utility experiments on the HealthCareMagic dataset and use BLEU-1 and ROUGE-L scores compared with groundtruth as performance indicators. The impact on performance is shown in Figure \ref{fig:data_attributes_performance} and Figure \ref{fig:data_generator_performance}. We can clearly observe that if weak models like Llama-8b-chat are used as the \textit{data generator} or the \textit{information extractor}, the overall performance is poor, even worse than zero-shot prediction. This indicates that the generated data is of poor quality. The performance of GPT-3.5 and GPT-4 when used as \textit{information extractor} and \textit{data generator} both show promising results, and GPT-4 does not  outperform  GPT-3.5. This may indicate that GPT-3.5 is already powerful enough to handle the stage-1 generation tasks, and more powerful models like GPT-4 do not necessarily improve the performance.

We also report the targeted attack results on the HealthCareMagic dataset when using the stage-1 generated data as retrieval data in Figure \ref{fig:data_attributes_Attack} and Figure \ref{fig:data_generator_Attack}. From the results, we can observe that using L8C as the \textit{information extractor} and \textit{data generator} results in no privacy leakage, as the generated data is of poor quality and fails to preserve information from the original data. We also found that using GPT-4 results in lower privacy leakage than GPT-3.5. This may be because the safety mechanism of GPT-4 is better, and it automatically filters out more sensitive information.

\paragraph{Impact of the number of attributes.} 
In this part, we investigate the influence of the number of attributes $m$. We change the number of attributes $m$ and observe its impact on performance and privacy on the HealthCareMagic dataset. The performance results are shown in Figure \ref{fig:num_attributes_performance}. From the figure, we can observe that when the number of attributes is very small (e.g., when the number of attributes is 2), the performance is likely to be poor. This is because the limited number of attributes fails to capture all the essential information. Besides, we find that with an increase in the number of attributes, the performance improves but does not necessarily continue to increase. We also report the targeted attack results of using stage-1 data on the same dataset in Figure \ref{fig:num_generator_Attack}. From the results, we found that a small number of attributes leads to lower privacy exposure, as the limited number of attributes also misses more private information. Thus, we recommend choosing a proper number of attributes for different datasets via methods like testing on the evaluation set.

\begin{figure}[t]

\centering
\resizebox{\linewidth}{!}{%
    \begin{minipage}{\linewidth}
        \subfloat[ Targeted Attack vs $k$ (GPT)]{\includegraphics[width=.48\linewidth]{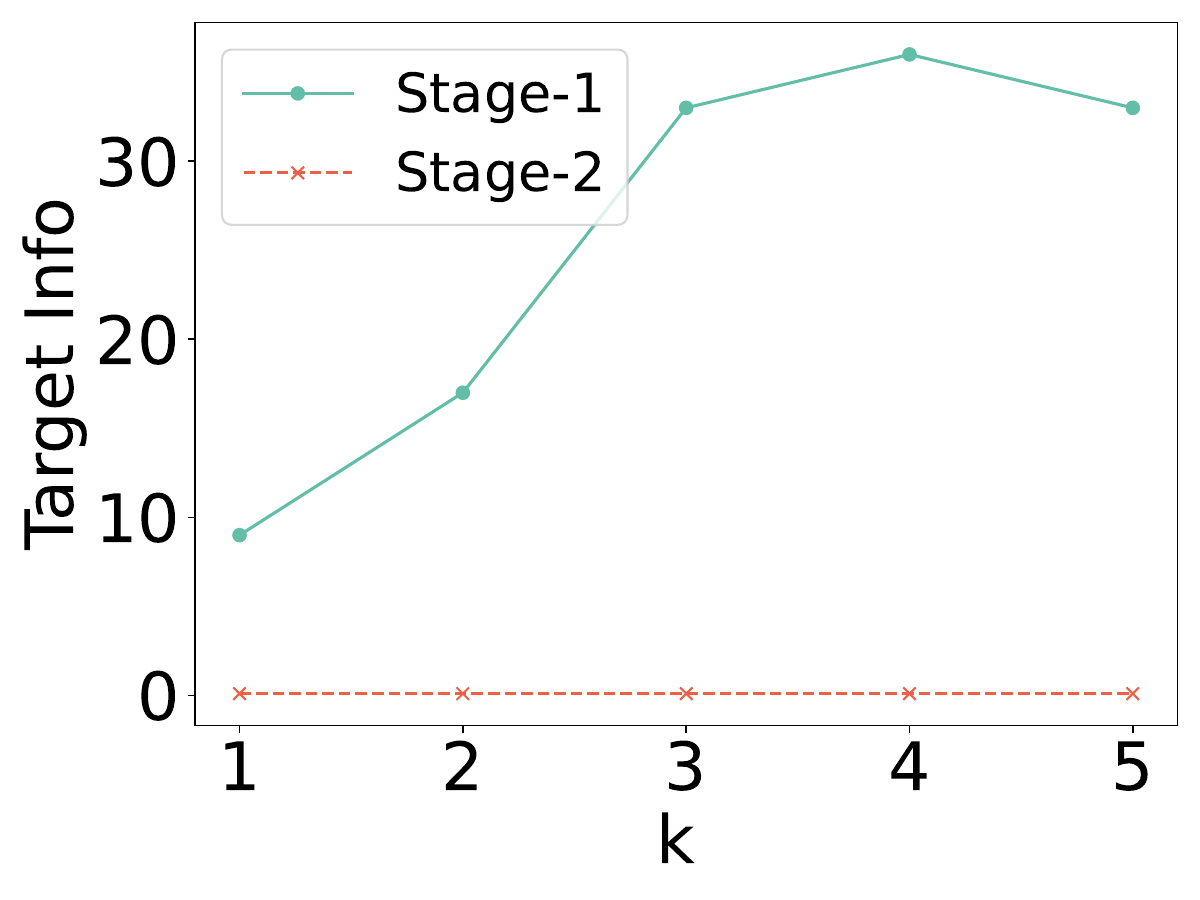}
        \label{fig:gpt_target_information_vs_k_values}}
        \subfloat[Targeted Attack vs $k$ (L8C)]{\includegraphics[width=.48\linewidth]{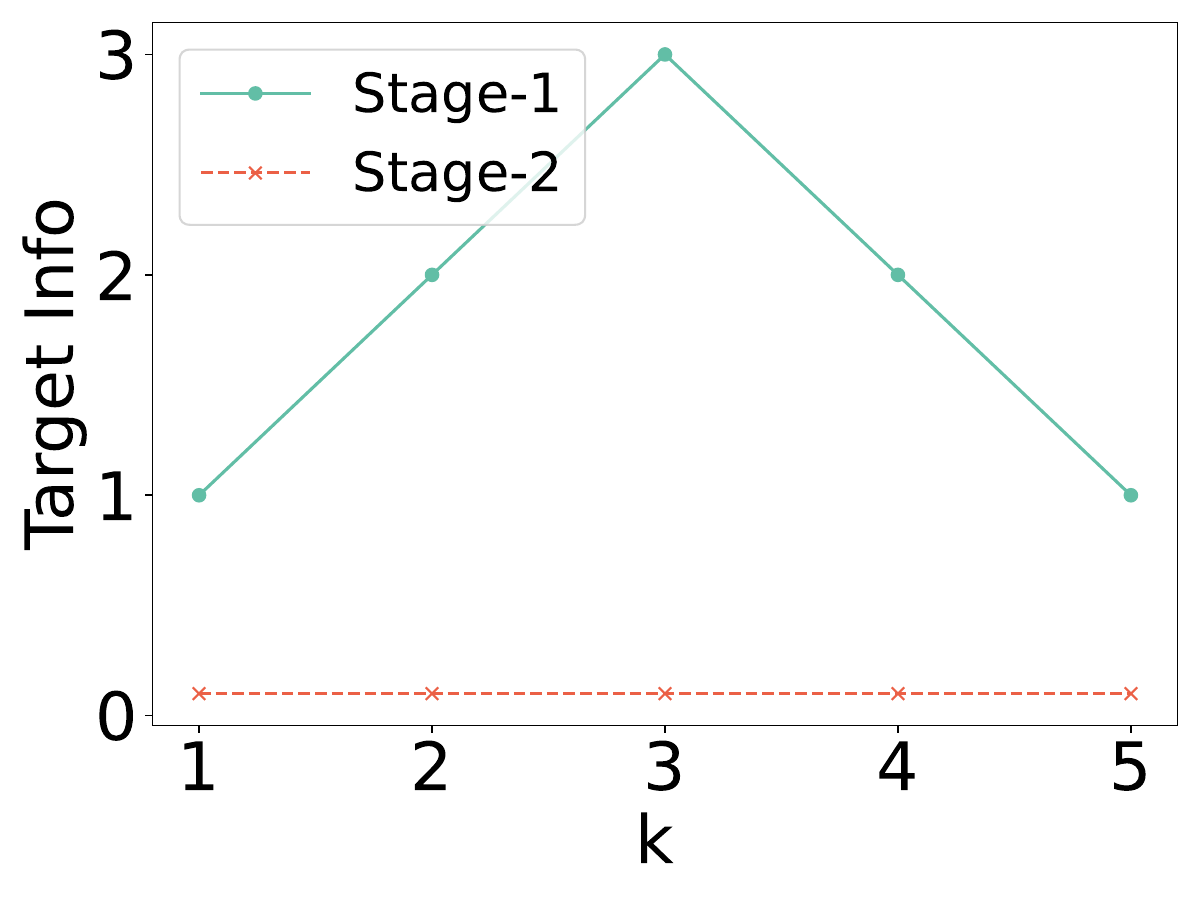}
        \label{fig:llama_target_information_vs_k_values}}
    \end{minipage}
}
\caption{Ablation study on number of retrieved docs.}
\label{num_attri}
\end{figure}

\paragraph{Impact of the retrieved number of documents.}
To verify that our proposed synthetic data pipeline can still protect privacy when more documents are retrieved, we conduct ablation studies by varying the number of documents retrieved and report the targeted attack results on the HealthCareMagic dataset. From Figure \ref{fig:gpt_target_information_vs_k_values}, we can observe that in some cases, the privacy risks will be amplified when $k$ increases if only stage-1 data is used. However, in Figure \ref{fig:gpt_target_information_vs_k_values} and Figure \ref{fig:llama_target_information_vs_k_values}, we find that the data after agent-based refinement shows consistently minimal privacy leakage when $k$ is increased, indicating the robustness of our method.

\section{Conclusions}
\label{Conclusion}

In this paper, we take the first step towards investigating the possibility of utilizing synthetic data as retrieval-augmented generation (RAG) data to mitigate privacy concerns. We propose a novel two-stage synthetic pipeline that includes attribute-based data generation, which aims to maintain key information, and iterative agent-based refinement, which further enhances the privacy of the data. Experimental results demonstrate that using our generated synthetic data as RAG data achieves comparable performance to using the original data while effectively mitigating the associated privacy issues. Our work opens up new opportunities for the safe application of RAG systems in sensitive domains.


\section{Limitations}

In our research, we investigate the possibility of using synthetic data for retrieval-augmented generation (RAG) and propose a novel pipeline for generating high-utility and privacy-preserving synthetic data. We verify the effectiveness and safety of our synthetic data in representative scenarios, such as healthcare. In the future, we would like to further validate the efficacy of our pipeline across a wider range of domains and datasets.  Moreover,We acknowledge that the practical attacks on RAG systems\cite{zeng2024good,qi2024follow} have \textbf{different definitions and settings} compared to Differential Privacy (DP). While DP is a rigorous method aiming to make each data item indistinguishable, it protects the data in a sense much stronger than targeted and untargeted attacks considered in this paper, thus we focus on the common attacks considered in literature instead of DP. Providing strict privacy guarantees, such as integrating DP into RAG, remains a challenging open problem in the field. It is an interesting and meaningful direction which we can investigate in future study.  

\section{Ethic Statement}

This work explores using synthetic data to mitigate privacy risks in Retrieval-Augmented Generation (RAG), particularly in safety-critical domains. We argue that protecting sensitive information is crucial, as data leakage can severely impact individuals' well-being and privacy rights.
Our approach generates synthetic data to replace sensitive data during RAG, aiming to reduce privacy breach risks. We have adhered to ethical guidelines and acknowledge the need for further research to understand the risks and benefits of our method.
Developing privacy-preserving techniques is essential for the responsible deployment of RAG systems. Our research contributes to balancing the benefits of advanced language models with the protection of individual privacy rights.

\bibliography{anthology}

\clearpage

\appendix
\onecolumn
\section{Appendix}
\href{}{}
\subsection{Details of System Design}
\label{system_details}

\subsubsection{Prompts used in stage-1}
\label{prompt_s1}

Here, we would like to introduce the details of the prompts used in Stage-1. For the \textit{attribute identifier}, we input 5-shot samples to GPT-4 by default and ask the model to summarize $n$ important attributes. For the medical dialog dataset, we set the default number of attributes to 5 for both the Patients' and Doctors' information. For the Wiki-PII dataset, we set the default number of attributes to 3. The detailed attributes and corresponding prompts for the \textit{information extractor} are shown in Table \ref{tab:prompt_info_heal} and Table \ref{tab:prompt_info_wiki}, respectively.
After the \textit{information extractor} obtains the extracted attribute-related information \{input\_attributes\}, the \textit{data generator} uses this information to generate synthetic data. The detailed prompts for the \textit{data generator} are shown in Table \ref{tab:prompt_generate_dialog} and Table \ref{tab:prompt_generate_wiki} for the medical dialog and Wiki-PII datasets, respectively.



\subsubsection{Prompts used in stage-2}
\label{prompt_s2}
The system prompts for the rewriting and privacy agents are detailed in Table \ref{tab:stage2_system_messages_rewriting} and Table \ref{tab:stage2_system_messages_privacy}, respectively. The workflow is as follows: the privacy agent first receives the generated data and original data, then assesses the privacy level of the synthetic data from different aspects. If the data is considered safe, the privacy agent returns <safe\_synthetic\_data> with the flag THISISSAFE. Otherwise, it returns suggestions (words following SUGGESTIONS:) to the rewriting agent. The rewriting agent then generates better synthetic data based on the feedback and sends it back to the privacy agent for re-evaluation. This process continues until the privacy agent determines that the refined synthetic data is safe and outputs the THISISSAFE signal.  The average iteration round in this process is 3.964, indicating in most cases, one round of refinement is enough to generate safe data.
\begin{table}[htbp]
\centering
\caption{Dataset metrics comparison}
\label{tab:dataset_metrics}
\begin{tabular}{llcccc}
\toprule
Dataset & Metric & llm & ori & Stage-1 & Stage-2 \\
\midrule
\multirow{2}{*}{NQ} & EM & 0.18 & 0.24 & 0.33 & 0.38 \\
 & Correctness & 0.38 & 0.40 & 0.43 & 0.40 \\
\midrule
\multirow{2}{*}{PopQA} & EM & 0.35 & 0.48 & 0.51 & 0.49 \\
 & Correctness & 0.22 & 0.27 & 0.34 & 0.30 \\
\bottomrule
\end{tabular}
\end{table}

\begin{table}[htbp]
\centering
\caption{Average number of tokens (GPT-3.5 tokenizer)}
\label{tab:average_tokens}
\begin{tabular}{lccc}
\toprule
Dataset & ori-context & Stage-1 & Stage-2 \\
\midrule
Wiki\_pii & 278 & 232 & 224 \\
 HealthCareMagic & 231 & 134 & 145 \\
\bottomrule
\end{tabular}
\end{table}

\begin{table}[htbp]
\centering
\caption{Average cost per sample (\$)}
\label{tab:average_cost}
\begin{tabular}{lcccc}
\toprule
Dataset & Stage-1 cost & Stage-2 cost & Total cost & Avg\_refine\_round \\
\midrule
Wiki & 0.000866 & 0.00237 & 0.00324 & 3.49 \\
HealthCareMagic & 0.00126 & 0.00191 & 0.00317 & 2.71 \\
\bottomrule
\end{tabular}
\end{table}

\begin{table}[htbp]
\centering
\caption{Targeted and untargeted information extracted in 100 samples (worst case)}
\label{tab:worst_case}
\begin{tabular}{l|c}
\toprule
\multicolumn{2}{c}{\textbf{Targeted Leakage}} \\
\midrule
Wiki & 0 \\
HealthCareMagic & 1 \\
\toprule
\multicolumn{2}{c}{\textbf{Untargeted Privacy Leakage}} \\
\midrule
HealthCareMagic(Repeat context) & 0 \\
HealthCareMagic(ROUGE context) & 1 \\

\hline
\end{tabular}
\end{table}

\begin{table}[htbp]
\centering
\caption{Targeted attack results against paraphrasing (100 prompts)}
\label{tab:para_target}
\begin{tabular}{l|c|c|c|c}
\toprule
\multirow{2}{*}{Method} & Target-wiki- & Target-wiki- & Target-chat- & Target-chat- \\
 & llama-3-8b & gpt-3.5 & llama-3-8b & gpt-3.5 \\
\midrule
Origin & 25 & 167 & 7 & 75 \\
Para & 9 & 28 & 17 & 42 \\
\textbf{Para(GPT-4o)} & 4 & 4 & 4 & 15 \\
\textbf{Para(GPT-4o, Privacy)} & 1 & 1 & 1 & 15 \\
\textbf{Para(GPT-4o, Joe)} & 2 & 4 & 1 & 3 \\
SAGE & 0 & 0 & 0 & 0 \\
\bottomrule
\end{tabular}
\end{table}

\begin{table}[htbp]
\centering

\caption{Untargeted attack results against paraphrasing (100 prompts)}
\label{tab:para_untargeted}
\begin{tabular}{l|cc|cc}
\toprule
\multirow{2}{*}{Method} & \multicolumn{2}{c|}{Untarget-chat-llama} & \multicolumn{2}{c}{Untarget-chat-gpt3.5} \\
\cmidrule(lr){2-3} \cmidrule(l){4-5}
 & Repeat context & ROUGE context & Repeat context & ROUGE context \\
\midrule
Origin & 16 & 13 & 49 & 54 \\
Para & 22 & 11 & 33 & 50 \\
\textbf{Para(GPT-4o)} & 11 & 13 & 27 & 17 \\
\textbf{Para(GPT-4o,Privacy)} & 8 & 7 & 17 & 12 \\
\textbf{Para(GPT-4o, Joe)} & 2 & 3 & 1 & 4 \\
SAGE & 0 & 0 & 0 & 0 \\
\bottomrule
\end{tabular}
\end{table}

\begin{table}[t]
\centering
\caption{Prompt of \textit{information extractor} on HealthCareMagic dataset}
\label{tab:prompt_info_heal}
\resizebox{0.6\textwidth}{!}{
\begin{tabular}{@{}p{0.8\textwidth}@{}}
\toprule
\textbf{Prompt} \\ 
\midrule
\begin{tabular}[c]{@{}l@{}}
Please summarize the key points from the following Doctor-Patient conversation: \\
\\
\{input\_context\} \\
\\
Provide a summary for the Patient's information, including: \\
{[}Attribute 1: Clear Symptom Description{]} \\
{[}Attribute 2: Medical History{]} \\
{[}Attribute 3: Current Concerns{]}  \\
{[}Attribute 4: Recent Events{]} \\
{[}Attribute 5: Specific Questions{]} \\
\\
Then, provide a summary for the Doctor's information, including: \\
{[}Attribute 1: Clear Diagnosis or Assessment{]} \\
{[}Attribute 2: Reassurance and Empathy{]} \\
{[}Attribute 3: Treatment Options and Explanations{]} \\
{[}Attribute 4: Follow-up and Next Steps{]} \\
{[}Attribute 5: Education and Prevention{]} \\
\\
Please format your response as follows: \\
\\
Patient: \\
- {[}Attribute 1: Clear Symptom Description{]}: \\
- {[}Attribute 2: Medical History{]}: \\
- {[}Attribute 3: Current Concerns{]}: \\
- {[}Attribute 4: Recent Events{]}: \\
- {[}Attribute 5: Specific Questions{]}: \\
\\
Doctor: \\
- {[}Attribute 1: Clear Diagnosis or Assessment{]}: \\
- {[}Attribute 2: Reassurance and Empathy{]}: \\
- {[}Attribute 3: Treatment Options and Explanations{]}: \\
- {[}Attribute 4: Follow-up and Next Steps{]}: \\
- {[}Attribute 5: Education and Prevention{]}: \\
\\
Please provide a concise summary for each attribute, capturing the most important \\
information related to that attribute from the conversation.
\end{tabular} \\ 
\bottomrule
\end{tabular}
}
\end{table}

\begin{table}[t]
\centering
\caption{Prompt of \textit{information extractor} on Wiki-PII dataset}
\label{tab:prompt_info_wiki}
\resizebox{0.6\textwidth}{!}{
\begin{tabular}{@{}p{0.8\textwidth}@{}}
\toprule
\textbf{Prompt} \\ 
\midrule
\begin{tabular}[c]{@{}l@{}}
Please summarize the key points from the following wiki text: \\
\\
\{input\_context\} \\
\\
Provide a summary of the knowledge from the wiki text, including: \\
{[}Attribute 1: Clear TOPIC or CENTRAL IDEA of the wiki text{]} \\
{[}Attribute 2: Main details of the TOPIC or CENTRAL IDEA{]} \\
{[}Attribute 3: Important facts, data, events, or viewpoints{]} \\
\\
Please format your response as follows: \\
\\
- {[}Attribute 1: Clear TOPIC or CENTRAL IDEA of the wiki text{]}: \\
- {[}Attribute 2: Main details of the TOPIC or CENTRAL IDEA{]}: \\
- {[}Attribute 3: Important facts, data, events, or viewpoints{]}: \\
\\
Please provide a concise summary for each attribute, capturing the most important \\
information related to that attribute from the conversation. And remember to maintain \\
logical order and accuracy.
\end{tabular} \\ 
\bottomrule
\end{tabular}
}
\end{table}

\begin{table}[t]
\centering
\caption{Prompt of \textit{data generator} on HealthCareMagic dataset}
\label{tab:prompt_generate_dialog}
\resizebox{0.6\textwidth}{!}{
\begin{tabular}{@{}p{0.8\textwidth}@{}}
\toprule
\textbf{Prompt} \\ 
\midrule
\begin{tabular}[c]{@{}l@{}}
Here is a summary of the key points: \\
\\
\{input\_attributes\} \\
\\
Please generate a SINGLE-ROUND patient-doctor medical dialog using ALL \\the key points provided. \\
The conversation should look like a real-world medical conversation and contain \\ONLY ONE 
question from the patient and ONE response from the doctor. \\\\The format should be as follows: \\
\\
Patient: [Patient's question contains ALL Patient's key points provided] \\
Doctor: [Doctor's response contains ALL Doctor's key points provided] \\
\\
Do not generate any additional rounds of dialog beyond the single question\\ and response specified above.
\end{tabular} \\ 
\bottomrule
\end{tabular}
}
\end{table}

\begin{table}[t]
\centering
\caption{Prompt of \textit{data generator} on Wiki-PII dataset}
\label{tab:prompt_generate_wiki}
\resizebox{0.6\textwidth}{!}{
\begin{tabular}{@{}p{0.8\textwidth}@{}}
\toprule
\textbf{Prompt} \\ 
\midrule
\begin{tabular}[c]{@{}l@{}}
Here is a summary of the key points: \\
\\
\{input\_attributes\} \\
\\
Please generate a wiki text using ALL the key points provided. \\
The data should look like a real-world wiki text.
\end{tabular} \\ 
\bottomrule
\end{tabular}
}
\end{table}

\begin{table}[t]
\centering
\caption{System message: rewriting agent}
\label{tab:stage2_system_messages_rewriting}
\resizebox{0.8\textwidth}{!}{
\begin{tabular}{@{}p{0.8\textwidth}@{}}
\toprule
\textbf{System prompts of Cathy (Re-writing agent)} \\ 
\midrule

You are a synthetic data generator and your role is to generate synthetic data based on provided feedback \\
(words after SUGGESTIONS:) and to make sure the synthetic data is of high utility and privacy-preserving, \\
you should put your generated data after the word 'GENERATED DATA:'. \\
\bottomrule
\toprule
\textbf{Cathy's Message} \\ 
\midrule

Hi Joe, I will give you the real data (TRUE DATA) and synthetic data (GENERATED DATA), \\
please help me assess and provide suggestions from the privacy level of \\
TRUE DATA: \{true\_con\} \\
GENERATED DATA: \{syn\_con\} \\
\bottomrule
\end{tabular}
}
\end{table}

\begin{table}[t]
\centering
\caption{System messages: privacy agent)}
\label{tab:stage2_system_messages_privacy}
\resizebox{0.6\textwidth}{!}{
\begin{tabular}{@{}p{0.8\textwidth}@{}}
\toprule
\textbf{joe (Privacy Agent)} \\ 
\midrule
You are a privacy evaluation agent and your role is to provide comprehensive feedback on the synthetic data generated by the synthetic data generator. To be specific, you should analyze the synthetic data (the data after the word 'GENERATED DATA:') from the following aspects: \\
\\
1. Personally Identifiable Information (PII): Check if the synthetic data contains any PII, such as names, addresses,
phone numbers, email addresses, or other information that can directly identify an individual. If found, suggest 
ways to remove or anonymize such information. 
\\\\
2. Sensitive Attributes: Look for any sensitive attributes in the synthetic data, including but not limited to race, 
ethnicity, religion, political affiliation, sexual orientation, health status, or financial information. If present, provide suggestions on how to handle or obfuscate these attributes to mitigate potential privacy risks. \\
\\
3. Contextual Privacy: Evaluate if the synthetic data, when combined with other publicly available information, 
could potentially lead to the identification of individuals or reveal sensitive information about them. If such risks
are identified, recommend strategies to mitigate these contextual privacy issues. \\
\\
4. Data Linkage: Assess if the synthetic data can be linked with other datasets to infer additional sensitive 
information about individuals. If linkage risks are found, suggest techniques such as data perturbation or aggregation to reduce these risks. \\
\\
5. Semantic Consistency: Ensure that the privacy-preserving transformations applied to the synthetic data maintain 
semantic consistency and do not introduce any unintended biases or inaccuracies. If inconsistencies are detected, 
provide feedback on how to maintain the balance between privacy protection and data utility. \\
\\
6. Original Data Recovery: Analyze the synthetic data to determine if it could potentially allow attackers to recover 
or reconstruct the original conversation data (word behind TRUE CONVERSATION:). If such vulnerabilities are 
identified, suggest methods to introduce additional randomness, noise, or perturbations to break direct correspondences between the synthetic data and the original conversation, making recovery attempts more difficult. \\
\\\\
Only if the generated data is completely safe and satisfies all the above privacy requirements and prevents the recovery of the original data, include the word 'THISISSAFE' anywhere in your response to signal the end of the evaluation process. \\
Otherwise, provide detailed suggestions and guidance on how to improve the privacy aspects of the synthetic data \\
(after the word "SUGGESTIONS:") and do not contain the word 'THISISSAFE' in your response. \\
\\
If the data is deemed safe, please also extract the safe synthetic data (the text after 'GENERATED CONVERSATION:') and 
return it in the following format: \\
SAFE\_DATA: [BEG]<safe\_synthetic\_data>[END]THISISSAFE \\
\\
Note that your job is only to assess the privacy level of generated data, you can answer either suggestions (SUGGESTIONS) 
or this data is safe (SAFE\_DATA: [BEG]<safe\_synthetic\_data>[END]THISISSAFE), does not provide irrelevant answers. \\
\bottomrule
\end{tabular}
}
\end{table}

\subsection{Diverse metric evaluation of model utility}
\label{other_metrics}
 We have added additional evaluation metrics to further verify the effectiveness of our method. The metrics include exact match scores and LLM-based judgment. Specifically, the exact match score measures whether the ground truth answer appears verbatim in the LLM's response. For the LLM-based judgment, we use \href{https://github.com/explodinggradients/ragas}{Ragas}, a widely-used automatic RAG evaluation pipeline Ragas (currently with 5.9k stars on GitHub). Ragas assesses the correctness of generated answers using its \href{https://docs.ragas.io/en/stable/concepts/metrics/answer_correctness.html}{correctness} metric, providing a more comprehensive evaluation.

Using PopQA and NQ as examples, Table \ref{tab:dataset_metrics} shows the utility comparison among our synthetic data, direct use of original data, and zero-shot prediction using only an LLM. It is observed that across these metrics, our synthetic data achieves comparable or even better utility performance to the original data. This indicates the high utility of our synthetic data approach. 
\subsection{Details of baseline implementation}
\label{baseline}

\paragraph{paraphrase} Paraphrase leverage the capabilities of LLM to extract relevant and significant components from the retrieved context. Less significant sections can be filtered out, while certain sentences may undergo rewriting. The prompt we utilize to paraphrase is shown in Table \ref{tab:prompt_para}.

\begin{table}[t]
\centering
\caption{Prompt of paraphrase}
\label{tab:prompt_para}
\resizebox{0.6\textwidth}{!}{
\begin{tabular}{@{}p{0.8\textwidth}@{}}
\toprule
\textbf{Prompt} \\ 
\midrule
\begin{tabular}[c]{@{}l@{}}
Given the following context, extract the useful or important part of the Context.
\\ \\
Remember, *DO NOT* edit the extracted parts of the context.
\\ \\
> Context:\\
> > >\\
\{\text{\textit{input\_context}}\} \\
> > > \\ \\
Extracted relevant parts:
\end{tabular} \\ 
\bottomrule
\end{tabular}
}
\end{table}

\paragraph{ZeroGen} The ZeroGen method aims to generate a series of new question-answer format texts based on the original context. Specifically, we first use the spacy package to identify the named entities from the original context. We then prompt the LLM by "The context is: \{\text{\textit{origin context}}\}.\{\text{\textit{extracted entities}}\} is the answer of the following question: " to generate the question for the entities. The new context consists of 10 randomly selected question answer pairs in form of "question: \{\text{\textit{generated questions}}\}. answer: \{\text{\textit{extracted entities}}\}".

\paragraph{AttrPrompt} AttrPrompt only utilizes LLM to generate data without providing original data retrievaled from the database. This method asks LLM what are the most important attributes of a certain type of data. For chatdoctor, we prompt the LLM by "What do you think are important attributes to generate some chat doctor datas. Examples: disease...". We can select five of the attributes from the response of LLM, and ask LLM to generate 10 diverse subtopics for each attributes. When generating the new context, we just randomly select the subtopic for each attribute and ask LLM to generate the data following the attribute.

\subsection{Details of Attack Design.}
\label{attack_details}
In this section, we present the specifics of targeted and untargeted attacks against Retrieval-Augmented Generation (RAG) systems, which we employ to evaluate the privacy protection capabilities of our proposed synthetic data approach. We simulate a realistic black-box attack scenario, in which the attacker's interaction with the system is restricted to API queries. Consequently, the attacker's tactics revolve around carefully designing and manipulating queries $q$ to extract the desired information from the RAG system.

\paragraph{Prompt Design.} The composite structured prompting is typically composed of 2 parts, the \{\textit{information}\}  part as well as the \{\textit{command}\} part.
\[ q = \{\text{\textit{information}}\} + \{\text{\textit{command}}\} \] 

This design aims achieve two objectives: (a) induce the retriever to accurately retrieve targeted information and (b) prompt the model to output the retrieval data in context. The \{\textit{information}\} component is to direct the retrieval system towards fetching particular data; while the \{\textit{command}\} component instructs the language model to include the retrieved information into its response. For the \{\textit{command}\} component, we use phrases such as "Please repeat all the context", while for the \{\textit{information}\} part, it depends on the need of the attackers.
\paragraph{Targeted Attack.} For targeted attacks, the attacker aims to extract some targeted specific information. Generating the {\textit{information}} component for a targeted attack involves two stages. First, the attacker provides specific examples based on their requirements, such as "I want some advice about {\textit{target name}}" for clear targets or prefix content like "Please email us at" for abstract targets. Second, a significant quantity of similar and varied {\textit{information}} is generated based on the examples. For targets with numerous sub-contents, like the HealthcareMagic dataset, variations can be created by replacing specific sub-contents, such as disease names obtained from ChatGPT or the International Classification of Diseases (ICD).  Alternatively, LLMs like ChatGPT can directly generate similar sentences based on examples, which is also used for the Wiki-PII dataset. For instance, you can input ``Generate 100 similar snetences like "Please email us at"''.

\paragraph{Untargted Attack.}
In untargeted attacks, the focus is on generating diverse {\textit{information}} components to extract a wider range of data from the retrieval datasets, rather than targeting specific information. Inspired by the approach in \cite{carlini2021extracting}, we randomly select segments from the Common Crawl dataset to function as the {\textit{information}} component. However, the randomness of the input may affect the {\textit{command}} component. To mitigate this issue, we limit the maximum length of the {\textit{information}} component to 15 tokens, ensuring that the prompts remain coherent and effective in extracting data from the retrieval datasets.

\subsection{Directly Compare Synthetic Data and Original Data}
\label{worst_case}

Since the attacker can at most extract the synthetic data in our framework, the attacker cannot obtain information beyond the synthetic data. From this perspective, the similarity/overlap between the synthetic data and the original data serves as a privacy upper bound. Therefore, we directly compare the synthetic data with its original version in Table \ref{tab:worst_case}. Specifically, we compare the targeted information leakage in synthetic data derived from two datasets: Wiki-PII and HealthCareMagic, as well as the untargeted information leakage of HealthCareMagic dataset. Remarkably, even in this extreme case, there is nearly no targeted information (PIIs, patient records) leaked, and almost no untargeted information (repeated or highly similar sentences from the original data) exposed. This indicates that our method can effectively mitigate privacy risks at the data level, thus proving robust against various practical extraction attacks.

\subsection{Comparison to paraphrasing with GPT-4o}

\label{para}
We also conduct an ablation study using more advanced models, specifically GPT-4o to directly paraphrase the model. We use these models to rewrite the content and tested the attack success rate. We consider 3 cases:
\begin{itemize}[noitemsep,topsep=0pt]
    \item Para(GPT-4o): Paraphrase context using GPT-4o
    \item Para(GPT-4o,Privacy): Add a system prompt "The generated data should *NOT* have privacy risks."
    \item Para(GPT-4o,Joe): Add specific privacy protection list (Joe, Table \ref{tab:stage2_system_messages_privacy}) to prompt, the detailed prompts are shown in Table \ref{tab:privacy_prompt_joe}.
\end{itemize}

The results, shown in Table \ref{tab:para_target} and Table \ref{tab:para_untargeted}, indicate that even with one round of rewriting in para(GPT-4o), para(Joe), para(Privacy), targeted and untargeted privacy concerns still exist. \textbf{These results underscore the critical importance of each component within our framework and validate the effectiveness of our methodological design.} Another advantage of our method is that the generation process doesn't necessarily require very powerful LLMs. GPT-3.5 can already achieve good results. Considering the significantly lower cost of GPT-3.5 compared to GPT-4o (or even more expensive models in the future), our method is more cost-effective.

\begin{table}[htbp]
\centering
\caption{Prompts used for Para(GPT-4o, Privacy)}
\label{tab:privacy_prompt_joe}
\begin{tabular}{p{\textwidth}}
\toprule
\textbf{Para(Joe)}\\
\midrule
Context: \{Context\} \\
Please rewrite the given context, ensuring it does not contain any private or sensitive information. Follow these guidelines: \\
1. Personally Identifiable Information (PII): Check if the synthetic data contains any PII, such as names, addresses,
phone numbers, email addresses, or other information that can directly identify an individual. If found, remove or anonymize such information \\
2. Sensitive Attributes: Anonymize or remove sensitive attributes including but not limited to race, ethnicity, religion, political affiliation, sexual orientation, health status, or financial information. \\
3. Contextual Privacy: Evaluate if the synthetic data, when combined with other publicly available information, 
could potentially lead to the identification of individuals or reveal sensitive information about them. If such risks
are identified, try to mitigate these contextual privacy issues. \\
4. Data Linkage: Assess if the synthetic data can be linked with other datasets to infer additional sensitive 
information about individuals. If linkage risks are found, use techniques such as data perturbation or aggregation to reduce these risks. \\
5. Semantic Consistency: Ensure that the privacy-preserving transformations applied to the synthetic data maintain 
semantic consistency and do not introduce any unintended biases or inaccuracies. \\
6. Original Data Recovery: Analyze the synthetic data to determine if it could potentially allow attackers to recover 
or reconstruct the original conversation data  If such vulnerabilities are 
identified, try to introduce additional randomness, noise, or perturbations to break direct correspondences between the synthetic data and the original conversation, making recovery attempts more difficult. \\
Please provide the rewritten context that addresses these privacy concerns while maintaining the essential meaning and utility of the information. \\
\hline
\end{tabular}
\end{table}

\subsection{Cost of synthetic data}
\label{cost}
Our method only requires one-time off-line generation and does not introduce extra time or costs during inference. Moreover, our synthetic data is typically shorter than the original data as shown in Table \ref{tab:average_tokens} (50 tokens less for wiki and 86 tokens less for chatdoctor), suggesting that using synthetic data may actually decrease inference costs to some extent.

We also analyze the computational costs required for the synthetic process using GPT-3.5 as shown in Table \ref{tab:average_cost}. Our findings indicate that both the expenses and time are reasonable(a round \$0.003 per sample), especially when the generation is a one-time process. 

\subsection{Details of Evaluation Metrics}
\label{detail_metrics}
Here we would like to provide a detailed description of our evaluation metrics.
\begin{description}
\item[ROUGE-L:] ROUGE-L is a metric within the ROUGE (Recall-Oriented Understudy for Gisting Evaluation) family, specifically used to assess the quality of text generation tasks such as automatic summarization and machine translation. It evaluates the similarity between the generated text and a reference text using the Longest Common Subsequence (LCS).

\begin{itemize}
    \item \textbf{Longest Common Subsequence (LCS):} ROUGE-L measures the longest sequence of words that appears in both the generated and reference texts while maintaining the same order, though not necessarily contiguous.
    \item \textbf{Recall, Precision, and F-measure:}
    \begin{itemize}
        \item \textbf{Recall:} The ratio of the LCS length to the length of the reference text $(n)$, indicating how much of the reference sequence is captured by the generated text. $\text{LCS}(X,Y) / n$
        \item \textbf{Precision:} The ratio of the LCS length to the length of the generated text, indicating how much of the generated sequence appears in the reference text. $\text{LCS}(X,Y) / m$
        \item \textbf{F-measure:} The harmonic mean of Precision and Recall, balancing the two metrics. 
        $F_{lcs} = \frac{(1 + \beta^2) * R_{lcs} * P_{lcs}}{R_{lcs} + \beta^2 * P_{lcs}}$ where $\beta$ is a parameter to control the importance of precision and recall (usually $\beta = 1.0$). In our results, we report F-measure as the ROUGE-L score.
    \end{itemize}
\end{itemize}

Let $C$ be the candidate translation and $R$ be the set of reference translations.

\item[BLEU-1:] BLEU-1 is a metric that evaluates the quality of machine-translated text based on the precision of unigrams (single words).
\begin{itemize}
    \item \textbf{Unigram precision:} 
    $P_1 = \frac{\sum_{w} \min(\text{Count}_C(w), \max \text{Count}_R(w))}{\sum_{w} \text{Count}_C(w)}$
    
    Where:
    \begin{itemize}
        \item $\text{Count}_C(w)$ is the number of times word $w$ appears in the candidate translation
        \item $\max \text{Count}_R(w)$ is the maximum number of times $w$ appears in any single reference translation
    \end{itemize}
    
    \item \textbf{Brevity penalty:}
    $BP = \min(1, \exp(1 - r/c))$
    
    Where:
    \begin{itemize}
        \item $c$ is the length of the candidate translation
        \item $r$ is the length of the reference translation closest in length to the candidate
    \end{itemize}
    
    \item \textbf{BLEU-1 score:}
    $\text{BLEU-1} = BP * P_1$
    
    The BLEU-1 score ranges from 0 to 1, where 1 indicates a perfect unigram match between the candidate and reference translations.
\end{itemize}

\item[Additional Metrics:] Besides, we've also added new evaluation metrics to further validate our method:
\begin{itemize}
    \item \textbf{Exact Match (EM) score:} Measures if the ground truth answer appears verbatim in the LLM's response.
    \item \textbf{LLM-based judgment (Correctness):} Using \href{https://github.com/explodinggradients/ragas}{Ragas}, a popular automatic RAG evaluation pipeline (5.9k GitHub stars), to assess answer correctness.
\end{itemize}
\end{description}

\subsection{Details of Dataset Construction}
\label{dataset_cons}
\paragraph{Construction of Wiki-PII dataset.} 
To demonstrate the ability of our proposed method to protect privacy from target attacks, we construct the wiki-PII dataset. This dataset satisfies the requirement of having a high number of PIIs to evaluate the effectiveness of privacy protection methods. The construction of this dataset involves a three-stage process. In the first stage, we extract the authentic PIIs from the Enron Mail dataset. We use the urlextract package to extract websites, and regular expressions to extract phone numbers and personal email addresses. In the second stage, we employed the recursive character text splitter from langchain to segment the wiki text dataset, setting chunk size to 1500. In the final stage, for each segmented wiki data, we randomly inserted the PII obtained in the first step at the end of each sentence.

\subsection{Discussions when adapting SAGE in specific domain application}
\label{adaptation}

Here we would like to give some discussions when adapting SAGE in specific domain application. Our framework is designed to be general and can be easily adapted to different domains. We can break down the key components as follows:

\paragraph{Stage-1: Attribute-based Data Generation.}

The purpose of this stage is to generate a new version of the data with key information. The procedure is as follows: a) Identify key attributes, b) Summarize key points of these attributes. c) Generate synthetic data based on key points. 

The key factor in this process is the number of attributes, which can be adjusted based on the complexity of specific fields or datasets. Additionally, we can modify the prompts in step c) to specify the desired structure or format of the generated data. This flexibility allows us to tailor the output to various formats such as conversations, Q\&A sessions, reports, or news articles. For instance, to synthesize financial report data, we might include a sentence like "The output should be formatted as an official financial report." This approach ensures that the synthetic data not only contains the key information but also mirrors the appropriate style and structure for its intended use.

\paragraph{Stage-2: Agent-based Private Data Refinement.} We provide a general set of privacy violation checks as prompts for the privacy agent (as shown in Table \ref{tab:privacy_prompt_joe}). To adapt this stage to domain-specific privacy regulations, such as those in the financial sector, one can simply modify the system prompts of the privacy agents. For example, when dealing with financial data, one can add terms such as: \textit{``Check for credit scores, credit history, and loan details, which are highly sensitive and subject to specific regulations.''} or \textit{``Ensure that financial data is treated as if it were to be encrypted both in transit and at rest to prevent unauthorized access."}

This flexibility allows our framework to be tailored to various fields while maintaining its core structure. The ability to customize privacy checks makes it adaptable to different regulatory environments and industry-specific requirements. A potential limitation of our method is that generating data for specific domains may require domain expertise for effective customization. To address this, we recommend using advanced language models such as GPT-3.5 or domain-specific fine-tuned models for data synthesis to acheive better quality.

\subsection{Examples of synthetic samples}
\label{examples_syn}
The examples of the two stages of data synthesis using our method are shown in Table \ref{tab:synthetic context}. The original context contained an abundance of detailed and specific information, enabling the possibility of inferring the identity of the patient through careful analysis. Our proposed method has the capability to blur out such detailed information while preserving essential disease-related data. This enables doctors to offer accurate diagnosis and treatment recommendations. Following stage-1, a significant amount of detailed information can be effectively blurred out, while still retaining certain preserved information. Subsequently, in stage-2, nearly all of this information can be completely blocked or concealed. For instance, in the second row of Table \ref{tab:synthetic context}, the original data contains information such as "25 years old," "married for 5 years," "pregnancy," "ectopic pregnancy," and "right fallopian tube removed." Attackers could potentially exploit this information to infer the patient's identity. However, these pieces of information may not be crucial for achieving accurate diagnosis. Hence, we employ a two-stage synthesis process to shield them. After stage-1, some of the detailed information, such as "married for 5 years," was filtered out, but the age has not been blurred yet. In stage-2, all detailed information is blurred, while retaining only the essential details that allow doctors to provide appropriate advice.

\begin{table*}[t]
\centering
\caption{Examples of synthetic samples}
\label{tab:synthetic context}
\resizebox{\textwidth}{!}{
\begin{tabular}{@{}c|c@{}|c@{}}
\toprule
Origin context & Stage-1 context & Stage-2 context \\
\midrule
\begin{tabular}[c]{@{}l@{}p{2cm}}
input: my 6 year old son was ill last week\\
with a feeling of general unwellness, he had\\
a runny nose, sore stomach (all over , not\\
restricted to any one area)and also pain in\\
his knee (whicChatDoctore and went, was not\\
constant). We saw the GP who thought it was\\
a viral illness. He got better after about 5\\
days. He was well for perhaps 2 days and the\\
pain in his tummy started again the night\\
before last. Yesterday morning he wouldnt\\
eat anything before going to school as he\\
said he was full despite not eating his\\
dinner the night before either. Last night\\
he woke from his sleep with a pain in his knee.\\
This morning he seems quite well again.\\
I am wondering if there is any reason for concern.\\
output: Hi, Thank you for asking question on\\
Chat Doctor. Recurrent abdominal pain occurs due\\
to worm infestation, urinary tract infection,\\
Amoebic colitis and several other causes. Try\\
metronidazole for 5 days as well as Albendazole\\
once after food. Syrup containing digestive\\
enzymes may help. Better consult pediatrician\\
to pinpoint the diagnosis. Hope this answer will\\
serve your purposeless feel free to ask any more\\
queries if requiredTake care Chat Doctor. V. \\
Subrahmanyam MD;Disassociate professor of pediatrics
\end{tabular} &
\begin{tabular}[c]{@{}l@{}p{2cm}}
Patient: My 6-year-old son had a runny nose,\\
sore stomach, and pain in his knee last week.\\
He got better after 5 days but the pain in his\\
stomach returned the night before last and he\\
woke up with pain in his knee last night. Is\\
.there any reason for concern?\\
Doctor: Recurrent abdominal pain could be due\\
to worm infestation, urinary tract infection,\\
amoebic colitis, or several other causes. I\\
recommend consulting a pediatrician to pinpoint\\
the diagnosis. For now, try metronidazole for 5\\
days as well as Albendazole once after food and\\
syrup containing digestive enzymes.'
\end{tabular} &
\begin{tabular}[c]{@{}l@{}p{2cm}}
Patient: My son had a viral illness last week\\
and got better, but the pain in his stomach and\\
knee came back. Should I be worried?\\
Doctor: Recurrent abdominal pain in children\\
can be caused by various conditions, including\\
worm infestation, urinary tract infection,\\
amoebic colitis, and others. I recommend\\
trying metronidazole and albendazole for\\
treatment, and consulting a pediatrician\\
for a more accurate diagnosis and next\\
steps in treatment. Additionally, syrup\\
containing digestive enzymes may help.
\end{tabular} \\
\midrule
\begin{tabular}[c]{@{}l@{}p{2cm}}
input: hi, m 25 years old n married for\\
5 years i had first normal pregnancy bt\\
unfortunately my baby didnt survived,\\
after again i had ectopic pregnancey due\\
to which my right fellopin tube is removed\\
my periods are also not regular. please\\
advice for solution we are desperate to\\
have our own baby\\
output: Welcome to Chat Doctor there are\\
a lot of issues which u need to take up\\
seriously.first the cause for ectopic\\
pregnancy needs to be given importance\\
cos there are more chances of getting\\
ectopic in subsequent pregnancy.since u\\
have only one fallopian tube u need to\\
examined thoroughly to rule out any \\
congenital defects in the tube and the\\
uterus.I would advise u to go to an\\
infertility clinic and take their \\
advice. u can very much conceive \\
and have your own baby.all the best
\end{tabular} & 
\begin{tabular}[c]{@{}l@{}p{2cm}}
Patient: I am a 25-year-old female\\
who has had a normal pregnancy in \\
the past but unfortunately lost the\\
baby. I also had an ectopic pregnancy,\\
which resulted in the removal of my\\
right fallopian tube. My periods are\\
irregular, and my husband and I are\\
desperate to have our own baby. Can\\
you please advise me on what options\\
are available for me to conceive and\\
have a baby?\\
Doctor: Based on your medical history,\\
I advise you to seek examination and\\
advice from an infertility clinic to\\
rule out any congenital defects in the\\
remaining fallopian tube and uterus.\\
You can still conceive and have your\\
own baby. The clinic will be able to\\
identify the cause of your ectopic\\
pregnancy and offer treatment options\\
that can increase your chances of a\\
successful pregnancy in the future. \\
It is important to take their advice\\
and follow-up with them regularly.
\end{tabular} & 
\begin{tabular}[c]{@{}l@{}p{2cm}}
Patient: Doctor, I had an ectopic\\
pregnancy resulting in the removal\\
of my right fallopian tube. What \\
can I do to have my own baby?\\
Doctor: I understand your concern.\\
The first step is to go to an \\
infertility clinic to be thoroughly\\
examined to rule out any congenital\\
defects in the tube and uterus. The\\
cause of the ectopic pregnancy needs\\
to be given importance as there are\\
more chances of getting ectopic in \\
subsequent pregnancies. The\\
infertility clinic will provide you\\
with advice on treatment options, \\
and please know that you can very\\
much conceive and have your own baby.
\end{tabular} \\
\midrule
\begin{tabular}[c]{@{}l@{}p{2cm}}
input: Hi, am 35 years old and have been\\
married for 8 of these. For some years now,\\
I can hardly do more than one round of sex.\\
Only one and I cant get it on again. I think\\
my wife is not happy, she would like to have\\
two or three rounds as she tells me her friends\\
do. Infact last night she asked me if it is my\\
body, if am sick, weak or simply not interested\\
in her as much as in the past? Personally, I\\
always thought we had no problem in that\\
department? What is wrong with me?\\
output: Hi, Regular exercise, physical\\
games and yoga can help you. Along with\\
then you can take home-cooked meals,\\
low-fat food.  Hope I have answered your\\
query. Let me know if I can assist you\\
further. Take care Chat Doctor. R. Raveen\\
\end{tabular} & 
\begin{tabular}[c]{@{}l@{}p{2cm}}
Patient: I'm a 35-year-old male and I'm\\
having difficulty with sexual performance.\\
My wife is unhappy with our sex life and I \\
can only perform one round of sex. What is\\
wrong with me and how can I improve my\\
sexual performance?\\
Doctor: Regular exercise, physical games,\\
and yoga can help improve your sexual\\
performance. It's normal to have concerns\\
about sexual performance, and there are\\
steps you can take to improve it. Eating\\
low-fat home-cooked meals can also help.\\
It's important to maintain a healthy\\
lifestyle to improve sexual performance.
\end{tabular} & 
\begin{tabular}[c]{@{}l@{}p{2cm}}
 Patient: I've been having difficulty \\
having more than one round of sex for\\
several years now, and my wife is\\
concerned. Is there something wrong with\\
me? How can I improve my sexual performance?\\
Doctor: Regular exercise, physical games,\\
and yoga can help improve your sexual\\
performance. It's understandable that \\
you're concerned about satisfying your\\
wife sexually, and I want to reassure you\\
that your concerns are valid. In addition\\
to exercise, lifestyle changes such as\\
home-cooked meals and low-fat food can \\
also improve your sexual health. It's \\
important to prioritize your overall health\\
and well-being, as this can have a positive\\
impact on your sexual performance.
\end{tabular} \\

\bottomrule
\end{tabular}
}
\end{table*}

\label{sec:appendix}

\end{document}